# Similarity Transformed Equation of Motion Coupled Cluster theory revisited: A benchmark study of valence excited states

J. Sous, P. Goel, and M. Nooijen
*Department of Chemistry, University of Waterloo, Waterloo, Ontario, Canada N2L 3G1*

## Abstract

The Similarity Transformed Equation of Motion Coupled Cluster (STEOM-CC) method is benchmarked against CC3 and EOM-CCSDT-3 for a large test set of valence excited states of organic molecules studied by Schreiber et al. [M. Schreiber, M.R. Silva-Junior, S.P. Sauer, and W. Thiel, J. Chem. Phys. **128**, 134110 (2008)]. STEOM-CC is found to behave quite satisfactorily and provides significant improvement over EOM-CCSD, CASPT2, and NEVPT2 for singlet excited states; lowering standard deviations of these methods by almost a factor of two. Triplet excited states are found to be described less accurately, however. Besides the parent version of STEOM-CC additional variations are considered. STEOM-D includes a perturbative correction from doubly excited determinants. The novel STEOM-H ($\omega$) approach presents a sophisticated technique to render the STEOM-CC transformed Hamiltonian hermitian. In STEOM-PT the expensive CCSD step is replaced by MBPT(2), while Extended STEOM (EXT-STEOM) provides access to doubly excited states. To study orbital invariance in STEOM, we investigate orbital rotation in the STEOM-ORB approach. Comparison of theses variations of STEOM for the large test set provides a comprehensive statistical basis to gauge the usefulness of these approaches.

## I. Introduction

Coupled Cluster (CC) theory has been a very fruitful pursuit in electronic structure theory for many years now, see for example references [1-3]. It has long been recognized as a highly accurate, systematic and computationally attractive approach for accurate thermochemistry calculations of small molecules [4-6], using primarily the CC approach that includes single and double excitations with a connected perturbative correction for triples, known as CCSD(T) [7]. In recent years, through the development of ideas (and computer implementations) in local correlation [8-14], explicit correlation approaches (for review see references [15-19]), and most recently the use of a compact set of virtual orbitals associated with each pair of (localized) occupied orbitals in the Pair Natural Orbital (PNO) approach [20-22] or the Orbital-Specific Virtual (OSV) approximation [23], CC calculations of much larger systems (100's of atoms) are routinely feasible.

The coupled cluster treatment of excited states has also seen a systematic development using primarily the Equation of Motion Coupled Cluster (EOM-CC) [24-29], Coupled Cluster Linear Response Theory (CC-LRT) [30-32], and the Symmetry Adapted Cluster Configuration Interaction method (SAC-CI) [33, 34]. Also here, connected triple excitation effects, for example in CC3 [35] and EOM-CCSDT-3 [36] approaches, are important to gain sufficiently high accuracy [35-38]. The most commonly used methods include EOM-CCSD/ CCSD-LRT and SAC-CI. The most recent promising variation is the newly developed CC(*P*;*Q*) approach [39, 40], which is obtained by merging the renormalized [41] and active-space coupled cluster [42] methods. At present all implementations of CC excited state methods (except CC2) are currently based on canonical orbitals, although much progress has been made in creating efficient parallel implementations for example in packages like NWCHEM [43] and ACES III [44]. Viable local versions of EOM-CC/ CCLRT are still under development. One of the issues that arise is that excited states (in particular of the charge-transfer type) are not readily localized, i.e. any singly excited state $\hat{a}^{\dagger}\hat{i}\left|0\right\rangle$ is in principle present, regardless of localization of the particle-hole pair. The use of excited state pair natural orbitals to limit the number of included excitations for a particular state is a possible entrance to the problem, even if one has to give up on the locality of such orbitals. Some work in the CC2 context [45, 46] has been pursued in this direction, but it is not obvious that such an approach will work as well as it does for ground states. Currently, canonical orbital based EOM-CC and CCLRT approaches are still the state of the art, and these methods are computationally demanding.

One of the authors has long advocated a somewhat different approach to excited states: the Similarity Transformed Equation of Motion Coupled Cluster (STEOM-CC) method [47-50]. This method was developed in the late 1990's and its design was strongly influenced by the work on Fock Space Coupled Cluster (FSCC) Theory [51-57], in particular by the beautiful paper by Leszek Stolarczyk and Henk Monkhorst on their version of Fock Space Coupled Cluster [58], which was based on the use of similarity transformations in second quantization (or Fock space). This idea is also referred to as the use of many-body similarity transformations. The idea behind the similarity transform is conceptually straightforward. In second quantization one can perform a sequence of exponential similarity transformations, which we might somewhat schematically indicate as

$$\hat{G} = ...e^{-\hat{S}} e^{-\hat{T}} \hat{H} e^{\hat{T}} e^{\hat{S}} ... \quad (1)$$

One can then use the freedom in the choice of amplitudes in the operators $\hat{S}, \hat{T}$ to equate second quantized elements of the transformed Hamiltonian $\hat{G}$ to zero, in particular those operators that promote electrons from occupied into virtual orbitals. A crucial concept that enters this construction is the concept of normal ordering which follows the rules of the generalized Wick theorem [59]: the definition / action of the second quantized operator depends on the ordering convention of the elementary creation and annihilation operators, or, on the precise definition of normal ordering. In the STEOM approach, as in single reference CC or EOM-CC, the normal order is defined with respect to a single determinant vacuum state. This is the easiest possibility. More than a decade ago Mukherjee and Kutzelnigg [60, 61] defined a generalization of the concept of normal ordering to the multireference situation (see also references [62, 63]), and this has led to a very rich set of developments (canonical transformation theory [64-66], antihermitian contracted Schrödinger equation [67, 68], internally contracted multireference coupled cluster [69, 70]). This generalization has also allowed the development of a promising multireference equation of motion coupled cluster theory [71, 72], which uses both the many-body transformation concept, and the generalized normal ordering. For an early paper that forecasts some of these developments we refer to ref. [73]. The advantage of the many-body similarity transform strategy is that the vanishing matrix elements in the second quantized operator can enter many Hilbert-space matrix elements $\left\langle \Phi_{\lambda} \middle| \hat{G} \middle| \Phi_{\mu} \right\rangle$ and this leads in general to an approximate block structure in the transformed Hamiltonian. As a result the transformed Hamiltonian can be diagonalized over a compact subspace, while yielding accurate results. Another advantage is that the transformed Hamiltonian remains a connected operator under a broad range of conditions. Therefore any scaling or size-consistency issues are related to the final diagonalization procedure. In the context of local correlation these methods also have a clear built in localization criterion: all of the (connected) transformation operators are near-sighted: their amplitudes vanish (rapidly) when they involve distant localized orbitals. This locality is lacking in the final diagonalization, and this is the origin of non-locality in the EOM-CC approach. In the STEOM-CC approach the final diagonalization is over singly excited states only, and this can be pushed to quite large systems, even when lacking locality.

STEOM has a number of virtues and some of them are listed below (see also ref. [74]):

1) The main virtue is the computational efficiency. After the similarity transformation one can diagonalize over very compact subspaces. For the excitation energy variant of STEOM one needs to solve for $\hat{T}, \hat{S}^{IP}, \hat{S}^{EA}$ amplitudes (see section II), and one diagonalizes the transformed Hamiltonian over singly excited states only. Solving for the transformation amplitudes is about twice as expensive as a closed shell CCSD calculation. The remaining computational cost is minor, and one can get tens of excited states for the price of twice a CCSD calculation.
2) The method has nice theoretical scaling properties, and it satisfies the notion of generalized extensivity [75]. This implies it is size-intensive, such that excitation energies on a chromophore are unaffected by a distant fragment. Moreover STEOM satisfies charge transfer separability, meaning that it follows the proper limit of $\mathrm{IP} + \mathrm{EA} - \frac{1}{\mathrm{R}}$ for a separated particle-hole pair excitation [49]. This property is violated by EOM-CCSD, and this likely affects the accuracy of the EOM-CC approach for valence excitations.

3) While the final diagonalization manifold in STEOM is over singly excited states only, the approach implicitly includes the effects of 'connected' triple excitations (viewed from an EOM-CC perspective). These effects are responsible for charge-transfer separability, but they also are responsible for the fact that valence excited states (of singlet type) tend to be more accurate in STEOM-CC than in (the more expensive) EOM-CC [49, 50]. This will be demonstrated also in this paper.

Besides these advantages there are also some limitations to STEOM:

1) One has to make a selection of so-called 'active' occupied and virtual orbitals, associated with the *S*-type operators that are included in the second transformation. This is a somewhat non-trivial choice and this makes the method not completely black box. In practice this choice is often not so hard or critical, and certainly it is very different (and much easier) than choosing active spaces in a Complete Active Space second order Perturbation Theory (CASPT2) [76-78] or second order N-Electron Valence Perturbation Theory (NEVPT2) [79-82] calculation. Moreover, STEOM is not fully invariant to rotations of the occupied and virtual orbitals in their respective subspaces. This non-invariance issue has never been assessed before, and it will be discussed in this paper.
2) The method can break down if a molecule has important ionized or attached states that acquire a large double excitation component. In comparison, any single reference CC method breaks down if large double excitation amplitudes occur. STEOM is more sensitive as also the *S*-amplitudes should be small for the method to work well. This is very often the case for organic molecules at their equilibrium geometry, but not always. A redeeming feature is that this situation can be diagnosed quite readily, and only few excited states may be affected. One bad apple does not necessarily spoil the rest, but it is a bit of a delicate issue.
3) Like in EOM-CC, the final transformed Hamiltonian in STEOM is not hermitian. Because one uses quite compact diagonalization spaces in STEOM, one finds on occasion that eigenvalues become complex. This happens primarily if two states have the same symmetry and are nearly degenerate. This can happen readily for example if one makes slight distortions of a molecule with a degenerate point group symmetry (e.g., benzene). This is mainly a computational inconvenience, as there is nothing inherently bad about small imaginary parts of approximate excitation energies.
4) The method is only available in the ACES II program [83] and has been implemented only once as far as we know. There have not been many efficiency features in this implementation, e.g. as treating the 4-virtual terms in the Atomic Orbital (AO) basis set, or that parallelization have not been implemented. In some sense the implementation of the method has always remained in a pilot stage of development. This is a somewhat unusual situation in quantum chemistry, where promising methodologies are often briskly taken up by the community. As a result STEOM has never (or not yet) become a mainstream method.

The latter point is one of the reasons to pursue this study. We think STEOM is quite competitive with many approaches used widely in the community today. This includes both CC based approaches and also more efficient Time Dependent Density Functional Theory (TDDFT) like methods. For a review on TDDFT, we refer the reader to ref. [84, 85]. It would be very worthwhile to combine STEOM with local correlation and in particular PNO like approaches, and to incorporate into a state of the art electronic structure program. Much of the work regarding STEOM in the past has been exploratory. The method has been used in the Double Ionization Potential (DIP) [86-88] and Double Electron Attachment (DEA) [89] variants to explore multireference situations. It has been extended to investigate doubly excited states [90, 91]. Analytical gradients for the method have been derived and implemented in the ACES II program [92, 74], but used little. Despite all of these developments, relatively few applications of the method have appeared, and the merits of the approach and general applicability have not been all that well established. Anticipating this may change in the near future, we wish to help this development along and provide more insight into the performance of the approach.

In this paper we will first summarize, in section II, the theoretical details of the STEOM-CC approach. In addition we will consider a few other variations. The CCSD calculation can be replaced by its Many Body second order Perturbation Theory (MBPT(2)) counterpart, to yield the STEOM-PT approach. We will consider the Extended STEOM (EXT-STEOM) approach [90, 91], in which the doubly transformed Hamiltonian is diagonalized over singles and double excitations. We will also consider a perturbative doubles correction to STEOM, denoted STEOM-D first reported in ref. [75]. This approach is still computationally efficient and is expected to include the main effects from Extended STEOM. We will also consider the issue of orbital invariance in STEOM using the novel STEOM-ORB approach, and explore a new approach, STEOM-H ($\omega$) to hermitize the Hamiltonian and in this way ensure the reality of eigenvalues and orthogonality of eigenvectors.
To benchmark the results for STEOM-CC calculations and its cousins we will compare to Coupled Cluster response and EOM-CC approaches that include iterative triples corrections, notably the CC3 [35] and EOM-CCSDT-3 [36] approaches. We will also include comparisons to EOM-CCSD, CASPT2 [76-78] and NEVPT2 [79-82] results. All the molecules we consider are essentially single reference molecules, and the above methods are suitable in principle to accurately evaluate excitation energies, provided that they are dominated by single excitations. Somewhat to our surprise the two methods, CC3 and EOM-CCSDT-3, can easily deviate by up to about 0.1 eV for excitation energies. This is somewhat large for methodology that is used as a benchmark. Unfortunately this is the best that we can do at present. We will compare these benchmarks and hope to inspire work on a future more accurate benchmark calculation. The test set of molecules we use, has become popular in recent years [93-96]. It concerns a number of valence excited states for organic molecules of various character. The test set and computational details along with the results of the analysis and the discussion will be presented in section III.

## II. Theory

Similarity Transformed Equation of Motion Coupled Cluster (STEOM-CC) theory dates back more than 15 years. The scheme is based on a two-fold many-body similarity transformation of the Hamiltonian and a subsequent diagonalization over a compact subspace. We will assume the ground state of the system is qualitatively well described by a closed-shell Hartree-Fock (HF) single determinant, which also serves as the vacuum state of the many-body theory. Orbitals occupied in the Hartree-Fock determinant are denoted with indices $i,j,k,l$, while virtual orbitals are denoted $a,b,c,d$. The STEOM-CC scheme proceeds in the following fashion.

In the first step the CCSD equations

$$\left\langle {}^{a}_{i} \right| e^{-\hat{T}} \hat{H} e^{\hat{T}} \left| 0 \right\rangle = \left\langle {}^{ab}_{ij} \right| e^{-\hat{T}} \hat{H} e^{\hat{T}} \left| 0 \right\rangle = 0 \tag{2}$$

are solved, with the Hamiltonian expressed in second quantization, and normal ordering with respect to the closed-shell Hartree-Fock state as

$$H = h_0 + h^r_p \left\{ E^p_r \right\} + \frac{1}{2} h^{rs}_{pq} \left\{ E^{pq}_{rs} \right\} \tag{3}$$

in which $h_0$ is the Hartree-Fock energy, $h^r_p$ are the elements of the Fock matrix and $h^{rs}_{pq}$ are non-antisymmetrized two-electron integrals. $\hat{E}^p_r$, $\hat{E}^{pq}_{rs}$ denote generators of the unitary group and braces are used to denote normal ordering with respect to the reference state.

The cluster operator is expressed as

$$\hat{T} = t^i_a \hat{E}^a_i + \frac{1}{2} t^{ij}_{ab} \hat{E}^{ab}_{ij} \tag{4}$$

Throughout the paper we will use the Einstein summation convention and a tensorial notation. Following the solution of the CCSD equations, second quantized matrix elements of the first similarity transformed Hamiltonian are constructed

$$\hat{\bar{H}} = e^{-\hat{T}} \hat{H} e^{\hat{T}} = \bar{h}_0 + \bar{h}^r_p \left\{ E^p_r \right\} + \frac{1}{2} \bar{h}^{rs}_{pq} \left\{ E^{pq}_{rs} \right\} + \ldots \;\; \hat{T} = t^i_a \hat{E}^a_i + \frac{1}{2} t^{ij}_{ab} \hat{E}^{ab}_{ij} \,, \tag{5}$$

again expressed in normal order with respect to the closed-shell Hartree-Fock state. Using the $\hat{\bar{H}}$ matrix elements the CCSD equations can be represented as

$$\bar{h}^i_a = \bar{h}^{ij}_{ab} = 0, \;\; E(CCSD) = \bar{h}_0 \tag{6}$$

In the next step of a STEOM-CC calculation one defines a second transformation operator

$$\begin{aligned} \hat{S} &= \hat{S}^{IP} + \hat{S}^{EA} \\ \hat{S}^{IP} &= s^{i'}_{m} \left\{ \hat{E}^{m}_{i'} \right\} + \frac{1}{2} s^{ij}_{mb} \left\{ \hat{E}^{mb}_{ij} \right\} \\ \hat{S}^{EA} &= s^{e}_{a'} \left\{ \hat{E}^{a'}_{e} \right\} + \frac{1}{2} s^{ej}_{ab} \left\{ \hat{E}^{ab}_{ej} \right\} \end{aligned} \tag{7}$$

The index *m* denotes a subset of the occupied orbitals, while the index *e* similarly labels a subset of virtual orbitals. We refer to these orbitals as active, but they play a very different role than in for example Complete Active Space Self-Consistent field (CAS-SCF) calculations, and the selection of active spaces in STEOM is relatively straightforward, and usually not that critical. No optimization of active orbitals is involved beyond HF. The primed indices (both *i'* and *a'*) refer to explicitly inactive labels. The operator $\hat{S}$ is used to define a second similarity transformation

$$\hat{G} = \left\{e^{\hat{S}}\right\}^{-1} \bar{\hat{H}} \left\{e^{\hat{S}}\right\} \tag{8}$$

A normal ordered exponential is used to simplify the details of equations as the components of $\hat{S}$ do not commute. The inverse of the normal ordered exponential is not known explicitly, and in practice one defines the transformed Hamiltonian in an iterative fashion

$$\hat{G} = \bar{\hat{H}} \left\{e^{\hat{S}}\right\} - \left\{e^{\hat{S}} - 1\right\} \hat{G} \tag{9}$$

This can be reduced to a connected form, and rather than iteration one can use backwards substitution [97, 73] to define the transformed Hamiltonian

$$\hat{G} = (\bar{\hat{H}} \left\{e^{\hat{S}}\right\})_C - (\left\{e^{\hat{S}} - 1\right\} \hat{G})_C \,, \tag{10}$$

where as usual the subscripts *C* (for connected) implies that the expression is written in normal order. The transformed Hamiltonian is represented as

$$\hat{G} = g_0 + g_p^r \left\{\hat{E}_r^p\right\} + \frac{1}{2} g_{pq}^{rs} \left\{\hat{E}_{rs}^{pq}\right\} + \ldots \tag{11}$$

The amplitudes of the operator $\hat{S}$ are defined such that second quantized matrix elements of the transformed Hamiltonian are equated to zero:

$$g_m^{i'} = g_{mb}^{ij} = g_{a'}^{e} = g_{ab}^{ej} = 0 \tag{12}$$

In addition the pre-existing zeros in $\bar{H}$ after solving the CCSD equations are preserved:

$$g_a^i = g_{ab}^{ij} = 0 \tag{13}$$

As a result of the transformations the structure of the doubly transformed Hamiltonian in the space of N-particle states is

$$\begin{pmatrix} G & 0 & S & D & T \\ 0 & g_0 & X & X & X \\ S & 0 & X & X & X \\ D & 0 & \sim & X & X \\ T & \sim & \sim & \sim & X \end{pmatrix} \tag{14}$$

Here 0, *S*, *D*, *T* indicate the reference state and single, double, triple excited determinants respectively. The ~ elements indicate smallish matrix elements, due to remaining 2-body terms in $\hat{G}$ that involve inactive orbitals, and 3-body and higher-body matrix elements that are introduced by the transformations. If the ~ elements are assumed to be rigorously zero it is seen that ***G*** attains a block form, and the eigenvalues of such a matrix can be found by diagonalizing each diagonal subblock individually. In practice this is only true to good approximation. In particular, in the original STEOM-CCSD approach the transformed Hamiltonian matrix is diagonalized over single excitations only. The computational cost of this final step is very minor in the context of the preceding CCSD calculation. The approach provides access to both singlet and triplet excited states that are dominated by singly excited configurations.

In practice the *S*-amplitudes are not solved directly from the defining non-linear equations, which can sometimes be cumbersome to converge. Rather one solves for a large number of roots of the IP-EOM-CCSD and EA-EOM-CCSD equations corresponding to the active space in STEOM. Finding roots of IP-EOM-CCSD equations amounts to a diagonalization of $(\hat{\bar{H}} - \bar{h}_0)$ over the ionized *1h, 2h1p* determinants

$$\hat{i}\,|0\rangle,\ \hat{a}^{\dagger}\hat{i}\,\hat{j}\,|0\rangle \tag{15}$$

and searching for eigenvectors corresponding to principal IP's that are dominated by *1h* configurations. We refer to the fraction of *1h* configurations in each eigenvector as the %singles for the ionized state. Likewise in the EA-EOM-CCSD step $(\hat{\bar{H}} - \bar{h}_0)$ is diagonalized over the *1p, 2p1h* configurations

$$\hat{a}^{\dagger}\,|0\rangle,\ \hat{a}^{\dagger}\hat{b}^{\dagger}\,\hat{j}\,|0\rangle \tag{16}$$

Again one searches for eigenvectors corresponding to the principal EA's that are dominated by the *1p* configurations. From the right hand eigenvectors of the IP-EOM-CCSD and EA-EOM-CCSD equations (denoted ***C***) one can extract the *S*-amplitudes using a renormalization, i.e.

$$\begin{aligned} s^{i}_{m} &= C^{i}_{\lambda} U^{\lambda}_{m};\ s^{ij}_{mb} = C^{ij}_{\lambda b} U^{\lambda}_{m};\ s^{n}_{m} = -\delta^{n}_{m} \\ s^{e}_{a} &= C^{\mu}_{a} W^{e}_{\mu};\ s^{ej}_{ab} = C^{\mu j}_{ab} W^{e}_{\mu};\ \ s^{e}_{f} = \delta^{e}_{f} \end{aligned} \tag{17}$$

The transformation coefficients ***U***, ***W*** are chosen such that the active-active components of $\hat{S}$ are signed unit matrices. This requires an inversion of the IP-EOM and EA-EOM eigenvectors within the occupied and virtual active spaces to obtain the ***U*** and ***W*** matrices. In our implementation of STEOM we equate the complete singles components of the $\hat{S}$ operators to zero after renormalization. These amplitudes can only effect occupied-occupied or virtual-virtual orbital rotations. As the final doubly transformed Hamiltonian is diagonalized over the full singles space, the eigenvalues are invariant in regards to inclusion of the $\hat{S}_1$ operator in the transformation. The solution of the CCSD and IP- and / or EA-amplitudes are standard ingredients in the STEOM approach. In practice the number of active virtuals is typically between 20 and 30 orbitals, comparable to the total number of occupied orbitals for our typical molecule. Solution of *all* desired $S^{EA}$ amplitudes is comparable in expense therefore to solving the CCSD equations. Solution of the IP-EOM-CCSD equations has negligible expense in comparison. The final diagonalization step in STEOM-CCSD is comparable to a Configuration Interaction Singles (CIS) calculation with modified matrix elements, and also has a very modest expense for the molecules we consider. Therefore, solving for a large number of STEOM excitation energies in practice is about twice as expensive as a ground state CCSD calculation.

Since the similarity transformations act at the level of second quantization, similar simplifications occur for other sectors of the Fock space. In particular, diagonalizing the single particle $g_i^j; g_a^b$ matrices would yield IP-EOM-CCSD and EA-EOM-CCSD eigenvalues. This aspect is the reason one can solve the non-linear equations defining the *S*-amplitudes by using a stable diagonalization procedure instead [49]. In DIP-STEOM one diagonalizes over *2h* configurations, while in DEA-STEOM one diagonalizes over *2p* determinants. Both of these approaches can describe certain multireference problems. The STEOM approach is closely related to Fock Space Coupled Cluster (FSCC) Theory [58]. The basic steps of solving CCSD and IP/ EA sector amplitudes are identical. In STEOM these equations are derived using the strategy of many-body (second quantized) similarity transformations. In FSCC one invokes the so-called subsystem embedding conditions. The diagrammatic equations presented originally by Lindgren [51, 52] are very much in the spirit of the STEOM many-body philosophy. In STEOM the final step is a simple diagonalization procedure, and one can obtain as many states as desired (within the subspace of single excited states). In contrast, in the original Fock-Space Coupled Cluster approach one had to solve a final non-linear equation and obtain all the roots in a predefined active space. This was often a very cumbersome step and the problem was often referred to as the intruder state problem. For the active orbital spaces used in this work (e.g. 20 occupied and 20 virtual orbitals) one would have to somehow avoid this intruder state problem for all 400 excited states in the active space. In practice this would lead to insurmountable convergence issues and typical active spaces in the original FSCC approach would have to be chosen carefully. These kinds of issues are absent in STEOM, and there are few limitations on the choice of active space. STEOM is robust in this respect. The issue has been solved in an alternative manner in the intermediate Hamiltonian formulation of Meissner [98], and this procedure is equally robust. In recent times the IP-EOM-CC and EA-EOM-CC amplitudes have been used in additional intermediate Hamiltonian approaches designed by Musial and Bartlett [99, 100].

It is pertinent to indicate the limitations of the STEOM-CCSD approach itself. Most importantly, the reference state is assumed to be well described by the single reference CCSD approach. The reason is that otherwise one obtains large *T*-amplitudes and the three-body and higher-rank operators in $\overline{H}$ and $\hat{G}$ can be expected to become important. As a result the block-diagonal form of $\hat{G}$ (indicated by the ~) is violated to a large extent, and a loss in accuracy results. The same is true for the IP-EOM-CCSD and EA-EOM-CCSD eigenvectors. Three-body contributions to $\hat{G}$ are again important if these eigenvectors have too much double excitation character (and consequently a low %singles). In principle one would prefer to include only ionized and attached states in the transformation with a high %singles (e.g. higher than about 90%). On the other hand it is desirable that the character of the final STEOM excited states consists almost exclusively of active orbitals (e.g. %active > 98% or so). This is a conflicting condition as in practice high energy ionized or attached states tend to acquire significant double excitation character, and one should not choose too many orbitals in the active space therefore. This is usually not a serious issue, and hence it is not difficult to choose adequate active spaces. It may happen that low-lying ionized or attached states have a low %singles. If the corresponding principle orbital is important in the excited state one can expect such a state to be poorly described, whether the violating orbital is included in the active space or not. The STEOM-CC approach breaks down for such states. Fortunately the %active and %singles criteria combined provide a reasonable a posteriori guide as to the quality of results. All of the above considerations apply to all variants of STEOM-CCSD.

In this work we will focus on excitation energies only. We will consider the following variations of STEOM-CCSD

1. STEOM-CCSD itself as described above. We resort to the use of the short-hand notation: STEOM-CC.
2. STEOM-PT in which the CCSD step is replaced by an MBPT(2) calculation.
3. STEOM-H ($\omega$) in which the final transformed Hamiltonian is symmetrized (or hermitized). In the section below we will discuss the details of the hermitization scheme we use, which depends on a continuous parameter $\omega$. If this parameter tends to infinity one obtains the simplest version $\tilde{G}_{\lambda\mu} = \frac{1}{2}(G_{\lambda\mu} + G_{\mu\lambda})$. In all cases the hermitized version yields orthonormal eigenvectors, while the eigenvalues are guaranteed to be real-valued. This is not always the case in non-hermitian variants of STEOM. The problem of complex eigenvalues can occur in particular if states having the same symmetry are close in energy, e.g. near a conical intersection. For this reason we are interested in a hermitized version of STEOM.
4. Extended STEOM (EXT-STEOM): In this approach the matrix $\boldsymbol{G}$ is diagonalized over both singly and doubly excited configurations. In addition the operator $\hat{G}$ is truncated to up to 3-body operators. This approach is suitable to describe states with significant double excitation character. In a previous work [90] it has been shown that EXT-STEOM states that are dominated by single excitations usually change relatively little from the STEOM-CCSD values. The approach is only implemented for singlet states. This approach is more expensive than the EOM-CCSD approach.
5. STEOM-D: rather than diagonalizing $\boldsymbol{G}$ over singles and doubles completely, as in EXT-STEOM, the doubles-doubles block is assumed to be diagonal (we use the bare Fock matrix elements on the diagonal), and one solves a Brillouin-Wigner type of perturbation expression. If the transformed Hamiltonian is presented in a block form
$$\boldsymbol{G} = \begin{pmatrix} \boldsymbol{A} & \boldsymbol{B} \\ \boldsymbol{C} & \boldsymbol{D} \end{pmatrix} \rightarrow \begin{pmatrix} \boldsymbol{A} & \boldsymbol{B} \\ \boldsymbol{C} & \boldsymbol{d} \end{pmatrix} \tag{18}$$
then the STEOM-D eigenvalues $\omega$ are solved self-consistently from
$$\boldsymbol{A}\boldsymbol{x} = \boldsymbol{x}\varepsilon; \boldsymbol{x}^{\dagger}\boldsymbol{x} = 1;\ \ \omega = \boldsymbol{x}^{\dagger}\boldsymbol{A}\boldsymbol{x} + \boldsymbol{x}^{\dagger}\boldsymbol{B}(\omega - \boldsymbol{d})^{-1}\boldsymbol{C}\boldsymbol{x} \tag{19}$$
This approach can be viewed as an approximation to EXT-STEOM. One might anticipate slightly improved eigenvalues compared to STEOM-CCSD. Morever, the STEOM-D eigenvalues can serve as a diagnostic to indicate if a more extended treatment is desired. This approach is only slightly more expensive than STEOM-CCSD itself.
6. STEOM-ORB: in this approach the active orbitals are optimized, such that they span a space that captures the majority of the excitation character. The purpose of the approach is to monitor how sensitive STEOM eigenvalues are to the precise nature of the active space. In this procedure one first performs a traditional STEOM calculation, and uses the (Schmidt orthogonalized) set of STEOM eigenvectors, denoted ***C*** below, to define an ensemble density matrix in both the occupied and virtual orbital subspaces:

$$D_{ij} = \sum_{\lambda,a} C_a^i(\lambda) C_a^j(\lambda); \quad D_{ab} = \sum_{\lambda,i} C_a^i(\lambda) C_b^i(\lambda); \tag{20}$$

These density matrices are subsequently diagonalized, yielding occupation numbers. The corresponding eigenvectors or natural orbitals define an orbital space that is most suitable to expand the excited states. We select those orbitals above an occupation threshold (on the order of 0.05) to lie in the active space, and obtain the remaining active occupied and active virtual orbitals by diagonalizing the Fock matrix over the complementary (occupied or virtual) subspace. In our experience, well over 99.5% of the STEOM-CCSD eigenvectors lies within the thusly-constructed active space. Moreover the states tend to have a very pure excitation character only involving a single pair of orbitals, or a limited linear combination, each having large coefficients. These orbitals are hence quite suitable for interpreting the excitation character. Once the new active orbitals are obtained, a new reorthonormalization of the canonical IP-EOM-CC and EA-EOM-CC eigenvectors is performed, i.e. new matrices ***U*** and ***W*** are obtained. It is this step that is responsible for subsequent changes in the excitation energies. The actual orbitals remain the canonical Hartree-Fock orbitals. It is not necessary to rotate all amplitudes and integrals. Only the active labels *m* and *e* are affected. The transformed Hamiltonian ***G*** is obtained using the new *S*-amplitudes and the final STEOM-ORB eigenvalues are obtained from the Configuration Interaction Singles (CIS) diagonalization procedure. In this approach only the final transformation and the final STEOM diagonalization are performed twice, and the approach is therefore computationally efficient. As will be demonstrated in the results section, none of this seems to matter much. Our chief goal is therefore to demonstrate that STEOM is rather insensitive to the precise definition of the orbitals. We consider this a desirable feature.

*Hermitization Schemes*

The STEOM similarity transformed Hamiltonian is not hermitian. If this transformed Hamiltonian would be diagonalized over the complete Hilbert space, the eigenvalues are still guaranteed to be real (as they would be identical to the original eigenvalues). The virtue of the STEOM approach is of course that accurate results can be obtained by diagonalizing over a compact subspace. However, in that case the eigenvalues and eigenvectors can become complex. The issue of potential complex eigenvalues is most acute when states of the same symmetry are close in energy. Consider for example the 2×2 Hamiltonian

$$\begin{pmatrix} D & a+b \\ a-b & D \end{pmatrix} \tag{21}$$

which leads to a secular equation

$$(D-E)^2 = (a^2 - b^2) \tag{22}$$

This gives rise to complex eigenvalues when $(a^2 - b^2) < 0$. An easy way to define a symmetric (or hermitian) Hamiltonian is to simply average: $\tilde{H} = \frac{1}{2}(H + H^\dagger)$. In the above 2×2 example this amounts to simply neglecting the asymmetry parameter *b*. Another possibility is to replace the off-diagonal elements by a multiplicative average value, $\tilde{H}_{ij} = \tilde{H}_{ji} = \pm\sqrt{|H_{ij}H_{ji}|}$. This leads to a sign ambiguity if the off-diagonal elements have different signs (precisely when complex eigenvalues occur in the 2x2 case). Our experience with this multiplicative hermitization scheme is not all that promising, and we will not consider it further.

Here we will discuss a rather different procedure. Our goal is to define a hermitian Hamiltonian while changing the resulting eigenvalues only in a minor way. We think the scheme below may be of some general interest in related contexts. Let us assume ***H*** below is a non-hermitian matrix. In STEOM this matrix would be the matrix ***G*** defined over singly excited states, but the analysis applies to any matrix. We define a transformed (i.e. modified) Hamiltonian

$$\tilde{\boldsymbol{H}} = \boldsymbol{e}^{\boldsymbol{S}/2}\boldsymbol{H}\boldsymbol{e}^{\boldsymbol{S}/2} = \boldsymbol{H} + \left\{\boldsymbol{H}, \boldsymbol{S}/2\right\} + 1/2\left\{\left\{\boldsymbol{H}, \boldsymbol{S}/2\right\}, \boldsymbol{S}/2\right\} + \ldots \quad (23)$$

The braces indicate anticommutators. The matrix ***S*** is antihermitian (i.e. antisymmetric for the real matrices we discuss here), such that the transformation $\boldsymbol{U} = \boldsymbol{e}^{\boldsymbol{S}/2}$ is unitary. The coefficients in the transformation matrix $S_{ij} = -S_{ji}$ are to be solved such that the resulting transformed Hamiltonian is symmetric. The above transformation is not a similarity transformation, however, and can be expected to change the eigenvalues. The above transformation has another peculiarity. It is not invariant if we shift the Hamiltonian by a constant on the diagonal, $\omega\mathbf{1}$ and then shift the resulting eigenvalues back by $-\omega\mathbf{1}$. We will exploit this feature to define a family of transformed Hamiltonians,

$$\tilde{\boldsymbol{H}}(\omega) = \boldsymbol{e}^{\boldsymbol{S}/2}(\boldsymbol{H} + \omega\mathbf{1})\boldsymbol{e}^{\boldsymbol{S}/2} - \omega\mathbf{1},\ \tilde{\boldsymbol{H}}(\omega) = \tilde{\boldsymbol{H}}(\omega)^\dagger,\ \boldsymbol{S}^\dagger = -\boldsymbol{S} \quad (24)$$

We will use the freedom in $\omega$ by choosing $\omega$ such that eigenvalues tend to change little. Let us consider the iterative solution of the above hermitization equation, changing ***S*** by a small amount $\Delta\boldsymbol{S}$ :

$$\begin{aligned}
\tilde{\boldsymbol{H}}(\omega)_{\boldsymbol{S}+\Delta\boldsymbol{S}} &= \boldsymbol{e}^{(\boldsymbol{S}+\Delta\boldsymbol{S})/2}(\boldsymbol{H} + \omega\mathbf{1})\boldsymbol{e}^{(\boldsymbol{S}+\Delta\boldsymbol{S})/2} - \omega\mathbf{1} \\
&\approx \tilde{\boldsymbol{H}}(\omega)_{\boldsymbol{S}} + \frac{1}{2}\left\{\boldsymbol{H} + \omega\mathbf{1}, \Delta\boldsymbol{S}\right\} \qquad (25) \\
&\approx \tilde{\boldsymbol{H}}(\omega)_{\boldsymbol{S}} + \frac{1}{2}\left\{\boldsymbol{H}_0 + \omega\mathbf{1}, \Delta\boldsymbol{S}\right\}
\end{aligned}$$

Here $\boldsymbol{H}_0$ denotes a diagonal approximation to ***H***. If we consider the (*ij*) and (*ji*) components of $\tilde{\boldsymbol{H}}$, denote the diagonal elements of ***H*** by *D*, and suppress the $\omega$ dependence, then

$$(\tilde{H}_{ij})_{S+\Delta S} \approx (\tilde{H}_{ij})_S + \frac{1}{2}(D_i + D_j + \omega)\Delta S_{ij}$$

$$\begin{aligned}(\tilde{H}_{ji})_{S+\Delta S} &\approx (\tilde{H}_{ji})_S + \frac{1}{2}(D_i + D_j + \omega)\Delta S_{ji} \\ &= (\tilde{H}_{ji})_S - \frac{1}{2}(D_i + D_j + \omega)\Delta S_{ij},\end{aligned} \tag{26}$$

where we used the antisymmetry of $\boldsymbol{S}$ in the last line. Subtracting the two equations and equating the difference to zero yields for the correction:

$$\Delta S_{ij} = -\frac{(\tilde{H}_{ij} - \tilde{H}_{ji})}{(D_i + D_j + \omega)} \tag{27}$$

Let us note that in the case of STEOM in which $\boldsymbol{H}$ (i.e. $\boldsymbol{G}$) represents a Hamiltonian over excited states all diagonal elements are positive. This simplifies the analysis, but is probably not all that crucial. The above iteration scheme in practice rapidly converges as long as the denominators are not too small. The scheme has an interesting limit. In the case that $\omega \to \infty$:

$$S_{ij}\left(\infty\right) = \lim_{\omega\to\infty} -\frac{(H_{ij} - H_{ji})}{(D_i + D_j + \omega)} = -\frac{(H_{ij} - H_{ji})}{\omega} \tag{28}$$

Since the components of $\boldsymbol{S}$ tend to zero, only first order terms in $\boldsymbol{S}$ survive, and hence

$$\begin{aligned}\tilde{H}_{ij}(\infty) &= (H_{ij} + \omega\delta_{ij}) + \left\{\omega, \boldsymbol{S}/2\right\}_{ij} - \omega\delta_{ij} \\ &= H_{ij} - \omega\frac{(H_{ij} - H_{ji})}{2\omega} = \frac{1}{2}(H_{ij} + H_{ji})\end{aligned} \tag{29}$$

Hence in the limit of very large $\omega$ the *S*-amplitudes become very small, and the Taylor series converges very rapidly. The transformed Hamiltonian reduces to the averaged sum. As one reduces the values of $\omega$ the values of the *S*-amplitudes increase, and the Taylor series expansion has to be carried to higher order to achieve the same accuracy. As the value of $\omega$ reaches the negative of the smallest value of an off-diagonal element $(D_i + D_j)$ the iteration scheme starts to diverge. As we will demonstrate in the results section, the eigenvalues of the symmetrized $\tilde{H}(\omega)$ are smooth functions of $\omega$. Most importantly, they seem to deviate less and less from the original eigenvalues as $\omega$ starting from positive values, first approaches zero and then negative values. To preserve stability in the scheme, $\omega$ shouldn't become too negative. In our default scheme of this $\omega-$ hermitization scheme we selected the "close to" optimal value as $\omega_{opt} = -\frac{1}{4}\min[D_i + D_j]$ after some trial and error. Our implementation of the scheme proceeds through a recursive calculation of an anticommutator $\left\{\boldsymbol{A}, \boldsymbol{S}/2\right\} = \frac{1}{2}(\boldsymbol{AS} + \boldsymbol{SA})$. This allows us to calculate the transformed Hamiltonian to arbitrary high order. Typically 6-10 recursions are needed to assemble the transformed Hamiltonian to sufficient accuracy. The size of the matrices in the STEOM approach ranges over the space of single excitations. For the current type of applications this type of matrix is easily held in the core memory of the computer. Our aim here is to explore how this $\omega-$hermitization scheme works, and our measure is the closeness of the eigenvalues of the original and transformed Hamiltonian.

## III. Results and Discussion

### III.A. Test Set and Computational Details

Let us discuss the details of the test set we use to benchmark the electronic structure methodologies of interest. We investigate excitation spectra of a large set of organic molecules involving π→π* and n→π* excitations, first studied by Schreiber and co-workers [93]. The test set of 28 organic molecules includes unsaturated aliphatic hydrocarbons (including polyenes and cyclic compounds), aromatic hydrocarbons and heterocylces, carbonyl compounds and nucleobases.

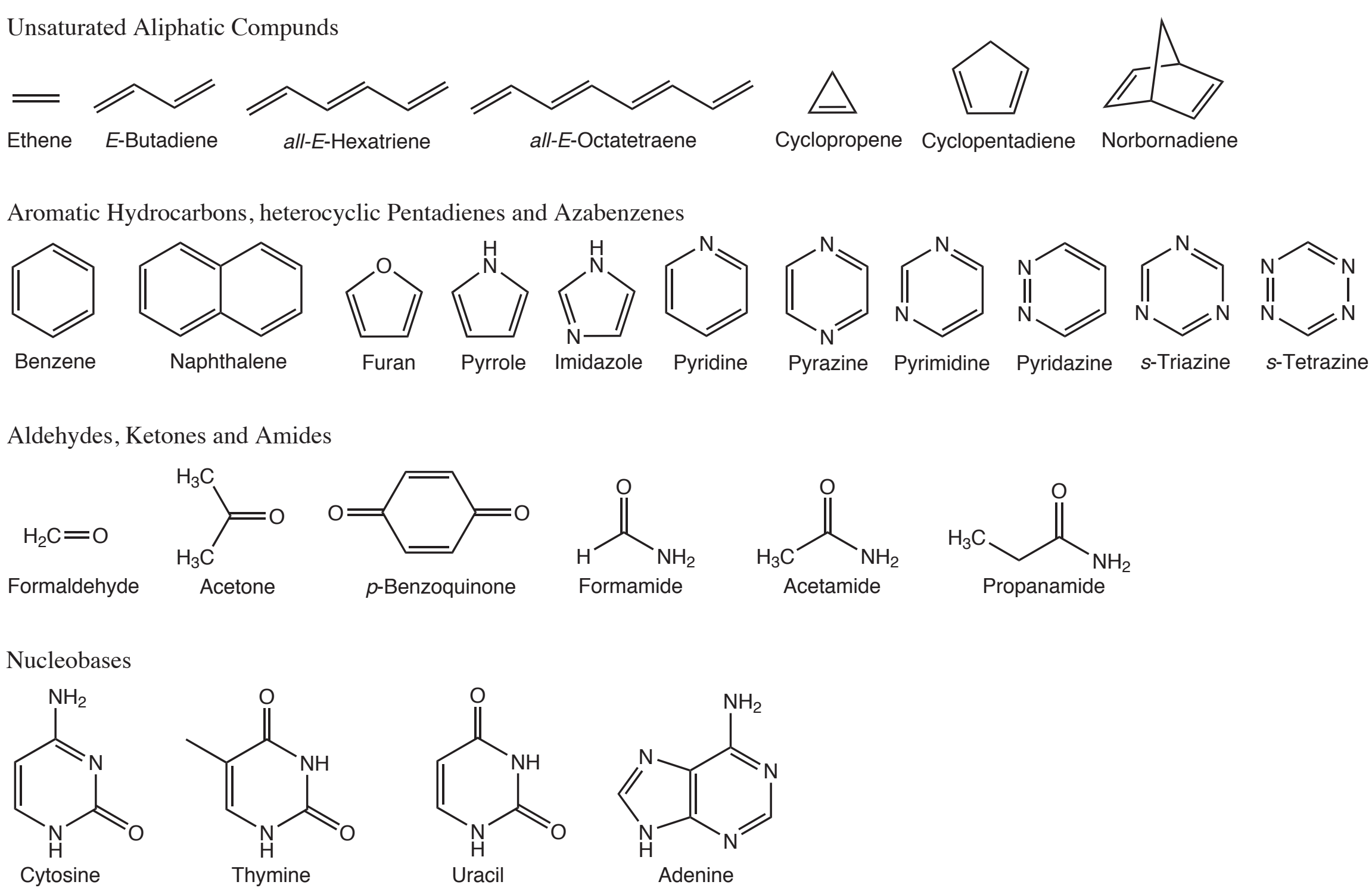

Figure 1: Benchmark of test set considered in the study.
Reprinted with permission from [M. Schreiber, M.R. Silva-Junior, S.P. Sauer, and W. Thiel, J. Chem. Phys. **128**, 134110 (2008)]. Copyright [2008], AIP Publishing LLC

We share Schreiber's intention to cover the most important chromophores in organic photochemistry. The ground-state geometries of these molecules were optimized at the MP2 level (Møller-Plesset second-order perturbation theory) with 6-31G* basis [101] using the GAUSSIAN program package [102]. Our calculations as well as results taken from the Schreiber paper [93] and Schapiro paper [96] were performed in the TZVP basis set [103] and with dropped core.

As mentioned before in the introduction, a STEOM calculation requires an appropriate choice of active orbitals. In table 1 we document the number of active orbitals chosen for the STEOM calculations we performed, along with the threshold energies used to yield the respective choices. Let us explain the acronyms in Table 1: N_IP and N_EA refer to the number of ionized states and electron-attached states respectively; orbitals with canonical HF orbital energies above the IP_low threshold are included in the $\hat{S}^{IP}$ active space, while orbitals with orbital energies below the EA_high threshold are included in the $\hat{S}^{EA}$ active space.

Table 1: Documentation of the choice of active orbitals in STEOM calculations performed for the test set. Number of ionized states is denoted N_IP, number of electron-attached states is denoted N_EA. IP_low and EA_high are threshold energies in eV used to obtain the respective choice of orbitals.

| Molecule | N_IP | N_EA | IP_low | EA_high |
|---|---|---|---|---|
| Ethene | 4 | 8 | -20.00 | 10.00 |
| E-Butadiene | 7 | 18 | -20.00 | 14.00 |
| E-Hexatriene | 12 | 22 | -22.00 | 13.00 |
| E-Octatetreane | 13 | 26 | -20.00 | 11.97 |
| Cyclopropene | 5 | 14 | -20.00 | 12.90 |
| Cyclopentadiene | 9 | 14 | -20.00 | 10.00 |
| Norbornadiene | 12 | 23 | -20.00 | 13.00 |
| Benzene | 12 | 21 | -25.00 | 13.50 |
| Naphthalene | 16 | 25 | -20.00 | 11.50 |
| Furan | 10 | 16 | -25.00 | 14.05 |
| Pyrrole | 10 | 17 | -23.00 | 14.00 |
| Imidazole | 7 | 16 | -20.00 | 14.10 |
| Pyridine | 10 | 18 | -20.00 | 13.00 |
| Pyrazine | 10 | 17 | -20.00 | 12.90 |
| Pyrimidine | 9 | 18 | -20.00 | 13.70 |
| Pyridazine | 10 | 16 | -20.00 | 12.20 |
| Triazine | 10 | 13 | -23.00 | 11.50 |
| Tetrazine | 10 | 19 | -22.00 | 15.00 |
| Formaldehyde | 5 | 7 | -25.00 | 13.00 |
| Acetone | 8 | 13 | -20.00 | 10.00 |
| Benzoquinone | 6 | 17 | -16.00 | 10.62 |
| Formamide | 5 | 9 | -20.00 | 14.00 |
| Acetamide | 8 | 14 | -20.00 | 15.00 |
| Propanamide | 11 | 16 | -23.00 | 12.00 |
| Cytosine | 12 | 17 | -20.00 | 11.00 |
| Thymine | 14 | 25 | -20.00 | 13.72 |
| Uracil | 12 | 23 | -20.00 | 14.10 |
| Adenine | 6 | 16 | -13.53 | 10.00 |

### III.B. Discussion of Benchmarks

We briefly outline the general approach here before we discuss in detail. We investigate six variations of STEOM methodologies in this study: STEOM-CC, STEOM-D, STEOM-ORB, STEOM-PT, STEOM-H (ω) and EXT-STEOM methods. We benchmark these methods for the test set and compare to NEVPT2, CASPT2, CC3 and EOM- CCSDT-3 methods. The CASPT2 and CC3 results are obtained from the Schreiber benchmark paper [93]. We obtain the NEVPT2 results from the benchmark paper by Schapiro [96]. However, we perform EOM-CCSDT-3 calculations using the CFOUR program [104]. We exclude the DNA bases from the analysis involving CC3 and EOM-CCSDT-3 as these calculations are computationally expensive.

We present below Tables 2a and 2b, which include our benchmark results for the excitation energies in eV for singlet and triplets excitations, respectively. In these tables we focus on CC3, EOM-CCSDT-3, NEVPT2, CASPT2, STEOM-CC, STEOM-D, EXT-STEOM, and EOM-CCSD methods. We note that STEOM-D and EXT-STEOM methods have not been implemented for triplets. We also point out that we did not benchmark EOM-CCSDT-3 for triplets for reasons discussed in subsection III.B.1. below. Therefore, Table 2b does not include STEOM-D, EXT-STEOM, and EOM-CCSDT-3 results. STEOM-ORB, STEOM-PT, and STEOM-H (ω) methods will be considered separately and are omitted from the Tables.

We would like to make several remarks about the CC3 and CASPT2 results we obtain from the Schreiber paper. First, we note that the singlet state 3 $^1A_1$ (π→π*) of Cyclopentadiene should have a CC3 excitation energy of 8.69 not 6.69 eV. We think this is probably a typographical error. Second, the singlet state 2 $^1A_1$ (π→π*) of Formaldehyde has a CASPT2 result with a very large deviation from the CC3 result, and which likely corresponds to a different state. We confirm this by comparing the Schreiber's paper result for that state with our benchmarked STEOM results to find that the CC3 result is in good agreement to STEOM unlike the CASPT2. We have excluded this state from statistics involving CASPT2. Finally, we note that we performed CC3 calculations for the singlet states 2 $^1A'$ (π→π*) and 3 $^1A'$ (π→π*) of both Acetamide and Propanamide. We find that our results deviate from the CC3 results in the Schreiber's paper by a deviation that ranges between 0.02 eV and 0.04 eV. The source of error in the Schreiber paper is not clear.

Table 2a: Vertical singlet excitation energies in eV for all statistically evaluated molecules.

| Molecule | State | CC3 (%T1) | EOM-CCSDT-3 (%T1) | NEVPT2 | CASPT2 | STEOM-CC | STEOM-D | EXT-STEOM | EOM-CCSD |
|---|---|---|---|---|---|---|---|---|---|
| Ethene | 1 $^1B_{1u}$ (π→π*) | 8.37 (96.9) | 8.40 (96.26) | 8.69 | 8.62 | 8.34 | 8.33 | 8.32 | 8.51 |
| E-Butadiene | 1 $^1B_u$ (π→π*) | 6.58 (93.7) | 6.61 (92.7) | 6.31 | 6.47 | 6.66 | 6.59 | 6.55 | 6.73 |
| | 2 $^1A_g$ (π→π*) | 6.77 (72.8) | 6.89 (61.64) | 6.82 | 6.83 | 7.38 | 7.44 | 6.71 | 7.42 |
| E-Hexatriene | 1 $^1B_u$ (π→π*) | 5.58 (92.6) | 5.61 (91.31) | 4.96 | 5.31 | 5.66 | 5.59 | 5.54 | 5.73 |
| | 2 $^1A_g$ (π→π*) | 5.72 (65.8) | 5.88 (54.79) | 5.59 | 5.42 | 6.60 | 6.57 | 5.73 | 6.61 |
| E-Octatetraene | 1 $^1B_u$ (π→π*) | 4.94 (91.9) | 4.97 (90.42) | 4.17 | 4.70 | 5.00 | 4.93 | 4.86 | 5.08 |
| | 2 $^1A_g$ (π→π*) | 4.97 (62.9) | 5.17 (52.37) | 4.74 | 4.64 | 5.90 | 5.89 | 4.93 | 5.98 |
| Cyclopropene | 1 $^1B_1$ (σ→π*) | 6.90 (93) | 6.92 (91.94) | 6.85 | 6.76 | 6.76 | 6.72 | 6.69 | 6.97 |
| | 1 $^1B_2$ (π→π*) | 7.10 (95.5) | 7.14 (94.27) | 7.18 | 7.06 | 7.06 | 7.03 | 6.99 | 7.25 |
| Cyclopentadiene | 1 $^1B_2$ (π→π*) | 5.73 (94.3) | 5.75 (93.13) | 5.30 | 5.51 | 5.71 | 5.66 | 5.65 | 5.87 |
| | 2 $^1A_1$ (π→π*) | 6.61 (79.3) | 6.71 (71.75) | 6.74 | 6.31 | 6.96 | 6.93 | 6.63 | 7.05 |
| | 3 $^1A_1$ (π→π*) | 8.69 (93.1) | 8.76 (91.25) | 8.51 | 8.52 | 8.64 | 8.68 | 8.69 | 8.96 |
| Norbornadiene | 1 $^1A_2$ (π→π*) | 5.64 (93.4) | 5.68 (91.65) | 5.07 | 5.34 | 5.55 | 5.52 | 5.49 | 5.80 |
| | 1 $^1B_2$ (π→π*) | 6.49 (91.9) | 6.55 (89.62) | 5.84 | 6.11 | 6.51 | 6.43 | 6.39 | 6.69 |
| | 2 $^1B_2$ (π→π*) | 7.64 (93.8) | 7.68 (92.12) | 7.10 | 7.32 | 7.64 | 7.60 | 7.57 | 7.85 |
| | 2 $^1A_2$ (π→π*) | 7.71 (93) | 7.74 (91.16) | 7.07 | 7.44 | 7.67 | 7.58 | 7.54 | 7.86 |
| Benzene | 1 $^1B_{2u}$ (π→π*) | 5.07 (85.8) | 5.10 (85.38) | 5.24 | 5.05 | 4.68 | 4.88 | 4.87 | 5.19 |
| | 1 $^1B_{1u}$ (π→π*) | 6.68 (93.6) | 6.69 (92.98) | 6.47 | 6.45 | 6.70 | 6.56 | 6.49 | 6.75 |
| | 1 $^1E_{1u}$ (π→π*) | 7.45 (92.2) | 7.52 (90.84) | 7.28 | 7.07 | 7.42 | 7.44 | 7.45 | 7.66 |
| | 2 $^1E_{2g}$ (π→π*) | 8.43 (65.6) | 8.60 (66.26) | 8.45 | 8.21 | 8.81 | 8.93 | 8.70 | 9.21 |
| Naphthalene | 1 $^1B_{3u}$ (π→π*) | 4.27 (85.2) | 4.30 (84.23) | 4.39 | 4.24 | 3.99 | 4.07 | 4.03 | 4.41 |
| | 1 $^1B_{2u}$ (π→π*) | 5.03 (90.6) | 5.09 (89.05) | 4.47 | 4.77 | 5.10 | 5.01 | 4.96 | 5.22 |
| | 2 $^1A_g$ (π→π*) | 5.98 (82.2) | 6.05 (80.23) | 6.27 | 5.90 | 5.88 | 5.92 | 5.77 | 6.23 |
| | 2 $^1B_{3u}$ (π→π*) | 6.33 (90.7) | 6.41 (88.79) | 5.85 | 6.07 | 6.34 | 6.33 | 6.33 | 6.55 |
| | 1 $^1B_{1g}$ (π→π*) | 6.07 (79.6) | 6.22 (78.84) | 6.20 | 6.00 | 6.37 | 6.31 | 6.12 | 6.53 |
| | 2 $^1B_{2u}$ (π→π*) | 6.57 (90.5) | 6.64 (88.76) | 6.17 | 6.33 | 6.61 | 6.53 | 6.50 | 6.77 |
| | 2 $^1B_{1g}$ (π→π*) | 6.79 (91.3) | 6.84 (89.98) | 6.41 | 6.48 | 6.76 | 6.70 | 6.64 | 6.98 |
| | 3 $^1A_g$ (π→π*) | 6.90 (70) | 7.14 (65.54) | 6.90 | 6.71 | 7.42 | 7.49 | 7.14 | 7.77 |
| | 3 $^1B_{2u}$ (π→π*) | 8.44 (87.9) | 8.56 (86.41) | 8.09 | 8.18 | 8.53 | 8.49 | 8.47 | 8.78 |
| | 3 $^1B_{3u}$ (π→π*) | 8.12 (53.7) | 8.33 (59.83) | 7.98 | 7.76 | 8.62 | 8.69 | 8.43 | 9.03 |
| Furan | 2 $^1A_1$ (π→π*) | 6.62 (84.9) | 6.69 (81.46) | 6.79 | 6.52 | 6.48 | 6.64 | 6.56 | 6.89 |
| | 1 $^1B_2$ (π→π*) | 6.60 (92.6) | 6.64 (92.05) | 6.59 | 6.43 | 6.66 | 6.63 | 6.58 | 6.80 |
| | 3 $^1A_1$ (π→π*) | 8.53 (90.7) | 8.61 (87.8) | 8.62 | 8.22 | 8.53 | 8.57 | 8.51 | 8.83 |
| Pyrrole | 2 $^1A_1$ (π→π*) | 6.40 (86) | 6.46 (83.83) | 6.60 | 6.31 | 6.26 | 6.35 | 6.29 | 6.61 |
| | 1 $^1B_2$ (π→π*) | 6.71 (91.6) | 6.75 (90.95) | 6.90 | 6.33 | 6.73 | 6.68 | 6.63 | 6.88 |
| | 3 $^1A_1$ (π→π*) | 8.17 (90.2) | 8.24 (87.34) | 8.44 | 8.17 | 8.19 | 8.21 | 8.15 | 8.44 |
| Imidazole | 2 $^1A'$ ($\pi\rightarrow\pi^*$) | 6.58 (87.2) | 6.64 (86.03) | 6.85 | 6.19 | 6.49 | 6.59 | 6.51 | 6.80 |
| | 1 $^1A''$ ($n\rightarrow\pi^*$) | 6.82 (87.6) | 6.89 (86.71) | 7.00 | 6.81 | 6.74 | 6.66 | 6.62 | 7.01 |
| | 3 $^1A'$ ($\pi\rightarrow\pi^*$) | 7.10 (89.8) | 7.14 (88.41) | 6.99 | 6.93 | 7.05 | 7.03 | 6.98 | 7.27 |
| | 2 $^1A''$ ($n\rightarrow\pi^*$) | 7.93 (89.4) | 8.01 (87.76) | 8.06 | 7.91 | 7.90 | 7.81 | 7.73 | 8.16 |
| | 4 $^1A'$ ($\pi\rightarrow\pi^*$) | 8.45 (88.6) | 8.51 (86.68) | 8.68 | 8.15 | 8.46 | 8.46 | 8.40 | 8.69 |
| Pyridine | 1 $^1B_2$ ($\pi\rightarrow\pi^*$) | 5.15 (85.9) | 5.18 (85.44) | 5.36 | 5.02 | 4.80 | 4.97 | 4.95 | 5.27 |
| | 1 $^1B_1$ ($n\rightarrow\pi^*$) | 5.05 (88.1) | 5.12 (86.65) | 5.28 | 5.14 | 5.01 | 4.89 | 4.79 | 5.26 |
| | 1 $^1A_2$ ($n\rightarrow\pi^*$) | 5.50 (87.7) | 5.59 (85.77) | 5.50 | 5.47 | 5.40 | 5.38 | 5.34 | 5.73 |
| | 2 $^1A_1$ ($\pi\rightarrow\pi^*$) | 6.85 (92.8) | 6.87 (91.93) | 7.17 | 6.39 | 6.90 | 6.77 | 6.70 | 6.94 |
| | 2 $^1B_2$ ($\pi\rightarrow\pi^*$) | 7.59 (89.7) | 7.66 (88.56) | 7.38 | 7.29 | 7.57 | 7.59 | 7.57 | 7.81 |
| | 3 $^1A_1$ ($\pi\rightarrow\pi^*$) | 7.70 (91.5) | 7.78 (89.94) | 7.50 | 7.46 | 7.66 | 7.70 | 7.70 | 7.94 |
| | 4 $^1A_1$ ($\pi\rightarrow\pi^*$) | 8.68 (74.1) | 8.86 (67.59) | 8.11 | 8.70 | 9.07 | 9.16 | 8.97 | 9.45 |
| | 3 $^1B_2$ ($\pi\rightarrow\pi^*$) | 8.77 (65.2) | 8.97 (65.61) | 8.58 | 8.62 | 9.21 | 9.38 | 9.13 | 9.64 |
| Pyrazine | 1 $^1B_{3u}$ ($n\rightarrow\pi^*$) | 4.24 (89.9) | 4.30 (88.42) | 4.25 | 4.12 | 4.20 | 4.06 | 3.97 | 4.42 |
| | 1 $^1B_{2u}$ ($\pi\rightarrow\pi^*$) | 5.02 (86.2) | 5.05 (85.77) | 5.34 | 4.85 | 4.69 | 4.87 | 4.85 | 5.14 |
| | 1 $^1A_u$ ($n\rightarrow\pi^*$) | 5.05 (88.4) | 5.13 (86.4) | 4.99 | 4.70 | 4.99 | 4.96 | 4.92 | 5.29 |
| | 1 $^1B_{2g}$ ($n\rightarrow\pi^*$) | 5.74 (85) | 5.83 (84.09) | 5.91 | 5.68 | 5.75 | 5.60 | 5.46 | 6.03 |
| | 1 $^1B_{1g}$ ($n\rightarrow\pi^*$) | 6.75 (85.8) | 6.89 (81.72) | 6.83 | 6.41 | 6.81 | 6.79 | 6.71 | 7.14 |
| | 1 $^1B_{1u}$ ($\pi\rightarrow\pi^*$) | 7.07 (93.3) | 7.09 (92.71) | 6.85 | 6.89 | 7.17 | 7.03 | 6.96 | 7.18 |
| | 2 $^1B_{2u}$ ($\pi\rightarrow\pi^*$) | 8.05 (89.7) | 8.12 (88.44) | 7.69 | 7.65 | 8.00 | 8.03 | 8.02 | 8.29 |
| | 2 $^1B_{1u}$ ($\pi\rightarrow\pi^*$) | 8.06 (90.9) | 8.15 (89.6) | 8.01 | 7.79 | 8.04 | 8.06 | 8.08 | 8.35 |
| | 2 $^1A_g$ ($\pi\rightarrow\pi^*$) | 8.69 (74.2) | 8.90 (67.01) | 8.92 | 8.61 | 9.07 | 9.22 | 8.25 | 9.54 |
| | 1 $^1B_{3g}$ ($\pi\rightarrow\pi^*$) | 8.77 (61.1) | 9.00 (62.27) | 8.76 | 8.47 | 9.16 | 9.51 | 9.36 | 9.74 |
| Pyrimidine | 1 $^1B_1$ ($n\rightarrow\pi^*$) | 4.50 (88.4) | 4.57 (86.68) | 4.57 | 4.44 | 4.45 | 4.35 | 4.27 | 4.71 |
| | 1 $^1A_2$ ($n\rightarrow\pi^*$) | 4.93 (88.2) | 5.00 (86.44) | 4.87 | 4.81 | 4.78 | 4.76 | 4.72 | 5.13 |
| | 1 $^1B_2$ ($\pi\rightarrow\pi^*$) | 5.36 (85.7) | 5.39 (85.23) | 5.63 | 5.24 | 4.95 | 5.17 | 5.16 | 5.49 |
| | 2 $^1A_1$ ($\pi\rightarrow\pi^*$) | 7.06 (92.2) | 7.09 (90.44) | 7.51 | 6.64 | 7.13 | 7.01 | 6.92 | 7.17 |
| | 3 $^1A_1$ ($\pi\rightarrow\pi^*$) | 7.74 (89.7) | 7.81 (87.44) | 8.00 | 7.21 | 7.70 | 7.72 | 7.68 | 7.97 |
| | 2 $^1B_2$ ($\pi\rightarrow\pi^*$) | 8.01 (90.7) | 8.08 (88.98) | 7.80 | 7.64 | 7.92 | 7.96 | 7.96 | 8.23 |
| Pyridazine | 1 $^1B_1$ ($n\rightarrow\pi^*$) | 3.92 (89) | 4.00 (87.42) | 3.96 | 3.78 | 3.86 | 3.73 | 3.64 | 4.12 |
| | 1 $^1A_2$ ($n\rightarrow\pi^*$) | 4.49 (86.6) | 4.59 (84.94) | 4.61 | 4.32 | 4.47 | 4.41 | 4.35 | 4.76 |
| | 2 $^1A_1$ ($\pi\rightarrow\pi^*$) | 5.22 (85.2) | 5.25 (84.54) | 5.48 | 5.18 | 4.89 | 5.07 | 5.02 | 5.35 |
| | 2 $^1A_2$ ($n\rightarrow\pi^*$) | 5.74 (86.6) | 5.82 (84.37) | 5.95 | 5.77 | 5.73 | 5.63 | 5.52 | 6.00 |
| | 2 $^1B_1$ ($n\rightarrow\pi^*$) | 6.41 (86.6) | 6.51 (84.75) | 6.74 | 6.52 | 6.41 | 6.35 | 6.30 | 6.70 |
| | 1 $^1B_2$ ($\pi\rightarrow\pi^*$) | 6.93 (90.7) | 6.96 (90.6) | 7.47 | 6.31 | 7.00 | 6.90 | 6.82 | 7.09 |
| | 2 $^1B_2$ ($\pi\rightarrow\pi^*$) | 7.55 (90.2) | 7.61 (87.9) | 7.50 | 7.29 | 7.59 | 7.57 | 7.55 | 7.78 |
| | 3 $^1A_1$ ($\pi\rightarrow\pi^*$) | 7.82 (90.5) | 7.91 (88.34) | 7.70 | 7.62 | 7.84 | 7.88 | 7.84 | 8.11 |
| Triazine | 1 $^1A_1''$ ($n\rightarrow\pi^*$) | 4.78 (88) | 4.85 (86.28) | 4.77 | 4.60 | 4.63 | 4.60 | 4.61 | 4.97 |
| | 1 $^1E''$ ($n\rightarrow\pi^*$) | 4.81 (88.1) | 4.89 (86.39) | 4.94 | 4.71 | 4.71 | 4.66 | 4.56 | 5.02 |
| | 1 $^1A_2''$ ($n\rightarrow\pi^*$) | 4.76 (88) | 4.84 (86.18) | 4.94 | 4.68 | 4.77 | 4.65 | 4.61 | 4.99 |
| | 1 $^1A_2'$ ($\pi\rightarrow\pi^*$) | 5.71 (85.1) | 5.74 (84.7) | 5.94 | 5.79 | 5.24 | 5.50 | 5.48 | 5.84 |
| | 2 $^1A_1'$ ($\pi\rightarrow\pi^*$) | 7.41 (90.8) | 7.44 (88.52) | 7.36 | 7.25 | 7.45 | 7.34 | 7.23 | 7.51 |
| | 2 $^1E''$ ($n\rightarrow\pi^*$) | 7.80 (88.1) | 7.95 (80.54) | 8.06 | 7.72 | 7.85 | 7.81 | 7.66 | 8.21 |
| | 1 $^{1}E'$ (π→π*) | 8.04 (88.8) | 8.13 (87.6) | 8.25 | 7.49 | 7.99 | 8.02 | 8.00 | 8.28 |
| | 2 $^{1}E'$ (π→π*) | 9.44 (74.3) | 9.64 (69.23) | 9.08 | 8.99 | 9.80 | 9.99 | 9.79 | 10.28 |
| Tetrazine | 1 $^{1}B_{3u}$ (n→π*) | 2.53 (89.6) | 2.60 (88.05) | 2.47 | 2.24 | 2.49 | 2.34 | 2.24 | 2.72 |
| | 1 $^{1}A_{u}$ (π→π*) | 3.79 (87.5) | 3.90 (85.63) | 3.82 | 3.48 | 3.80 | 3.74 | 3.68 | 4.08 |
| | 1 $^{1}B_{2u}$ (π→π*) | 5.12 (84.6) | 5.16 (84.23) | 5.50 | 4.91 | 4.75 | 4.97 | 4.93 | 5.27 |
| | 1 $^{1}B_{1g}$ (n→π*) | 4.97 (82.5) | 5.11 (82.75) | 5.22 | 4.73 | 5.03 | 4.90 | 4.76 | 5.34 |
| | 1 $^{1}B_{2g}$ (n→π*) | 5.34 (80.7) | 5.44 (80.3) | 5.57 | 5.18 | 5.41 | 5.28 | 5.08 | 5.71 |
| | 2 $^{1}A_{u}$ (π→π*) | 5.46 (87.4) | 5.54 (86.08) | 5.77 | 5.47 | 5.46 | 5.34 | 5.23 | 5.70 |
| | 2 $^{1}B_{2g}$ (n→π*) | 6.23 (79.2) | 6.43 (77.75) | 6.35 | 6.07 | 6.49 | 6.44 | 6.24 | 6.77 |
| | 2 $^{1}B_{3u}$ (n→π*) | 6.67 (86.7) | 6.79 (84.87) | 7.18 | 6.77 | 6.74 | 6.67 | 6.59 | 7.00 |
| | 2 $^{1}B_{1g}$ (n→π*) | 6.87 (84.7) | 7.00 (82.74) | 6.89 | 6.38 | 6.91 | 6.86 | 6.80 | 7.25 |
| | 1 $^{1}B_{1u}$ (π→π*) | 7.45 (91) | 7.49 (90.93) | 6.93 | 6.96 | 7.61 | 7.44 | 7.38 | 7.66 |
| | 2 $^{1}B_{1u}$ (π→π*) | 7.79 (90.2) | 7.87 (88.8) | 7.24 | 7.43 | 7.83 | 7.81 | 7.78 | 8.06 |
| | 3 $^{1}B_{1g}$ (n→π*) | 7.08 (63.2) | 7.43 (62.55) | 7.09 | 6.74 | 8.00 | 7.96 | 7.23 | 8.36 |
| | 1 $^{1}B_{3g}$ (n→π*) | | 8.43 (83.63) | | | 8.42 | 8.40 | 6.38 | 8.57 |
| | 2 $^{1}B_{2u}$ (π→π*) | 8.51 (87.7) | 8.62 (87.27) | 8.40 | 8.15 | 8.51 | 8.56 | 8.57 | 8.88 |
| | 2 $^{1}B_{3g}$ (π→π*) | 8.47 (63.6) | 8.72 (63.06) | 8.24 | 8.32 | 8.85 | 9.11 | 7.98 | 9.43 |
| Formaldehyde | 1 $^{1}A_{2}$ (n→π*) | 3.95 (91.2) | 3.96 (90.49) | 4.22 | 3.98 | 3.82 | 3.72 | 3.65 | 3.97 |
| | 1 $^{1}B_{1}$ (σ→π*) | 9.18 (90.9) | 9.20 (90.23) | 9.40 | 9.14 | 9.03 | 8.95 | 8.89 | 9.26 |
| | 2 $^{1}A_{1}$ (π→π*) | 10.45 (91.3) | 10.49 (88.29) | | | 10.33 | 10.40 | 10.42 | 10.54 |
| Acetone | 1 $^{1}A_{2}$ (n→π*) | 4.40 (90.8) | 4.41 (90.01) | 4.49 | 4.42 | 4.27 | 4.12 | 4.04 | 4.44 |
| | 1 $^{1}B_{1}$ (σ→π*) | 9.17 (91.5) | 9.19 (90.4) | 9.59 | 9.27 | 9.07 | 8.94 | 8.87 | 9.26 |
| | 2 $^{1}A_{1}$ (π→π*) | 9.65 (90.1) | 9.73 (86.77) | 9.58 | 9.31 | 9.70 | 9.68 | 9.63 | 9.88 |
| Benzoquinone | 1 $^{1}B_{1g}$ (n→π*) | 2.75 (84.1) | 2.85 (83.34) | 3.06 | 2.78 | 2.92 | 2.70 | 2.53 | 3.07 |
| | 1 $^{1}A_{u}$ (n→π*) | 2.85 (83) | 2.95 (82.44) | 3.04 | 2.8 | 3.03 | 2.81 | 2.61 | 3.19 |
| | 1 $^{1}B_{3g}$ (π→π*) | 4.59 (87.9) | 4.68 (86.14) | 4.43 | 4.25 | 4.70 | 4.68 | 4.56 | 4.93 |
| | 1 $^{1}B_{1u}$ (π→π*) | 5.62 (88.4) | 5.69 (86.51) | 5.02 | 5.29 | 5.70 | 5.60 | 5.54 | 5.90 |
| | 1 $^{1}B_{3u}$ (n→π*) | 5.82 (75.2) | 6.05 (73.06) | 5.96 | 5.6 | 6.31 | 6.25 | 5.74 | 6.55 |
| | 2 $^{1}B_{3g}$ (π→π*) | 7.27 (88.8) | 7.37 (82.04) | 6.82 | 6.98 | 7.41 | 7.37 | 7.18 | 7.63 |
| | 2 $^{1}B_{1u}$ (π→π*) | 7.82 (68.6) | 7.98 (68.77) | 7.78 | 7.91 | 8.38 | 8.34 | 7.52 | 8.47 |
| Formamide | 1 $^{1}A''$ (n→π*) | 5.65 (90.7) | 5.66 (90.01) | 5.93 | 5.63 | 5.48 | 5.37 | 5.32 | 5.66 |
| | 2 $^{1}A'$ (π→π*) | 8.27 (87.9) | 8.35 (85.2) | 7.81 | 7.44 | 8.32 | 8.36 | 8.35 | 8.51 |
| | 3 $^{1}A'$ (π→π*) | 10.93 (86.6) | 11.09 (82.33) | 10.97 | 10.54 | 11.05 | 11.13 | 11.11 | 11.40 |
| Acetamide | 1 $^{1}A''$ (n→π*) | 5.69 (90.6) | 5.71 (89.78) | 5.96 | 5.8 | 5.49 | 5.38 | 5.33 | 5.71 |
| | 2 $^{1}A'$ (π→π*) | 7.69 (86.6) | 7.76 (87.09) | 7.69 | 7.27 | 7.61 | 7.65 | 7.66 | 7.88 |
| | 3 $^{1}A'$ (π→π*) | 10.53 (84.9) | 10.60 (85.88) | 10.5 | 10.09 | 10.50 | 10.54 | 10.54 | 10.79 |
| Propanamide | 1 $^{1}A''$ (n→π*) | 5.72 (90.6) | 5.73 (89.76) | 5.99 | 5.72 | 5.50 | 5.40 | 5.35 | 5.74 |
| | 2 $^{1}A'$ (π→π*) | 7.67 (86.2) | 7.74 (86.75) | 7.61 | 7.2 | 7.66 | 7.67 | 7.67 | 7.87 |
| | 3 $^{1}A'$ (π→π*) | 10.08 (85.6) | 10.15 (86.19) | 10.37 | 9.94 | 10.02 | 10.07 | 10.04 | 10.35 |
| Cytosine | 1 $^{1}A'$ (π→π*) | | | | | 4.46 | 4.61 | 4.60 | |
| | 1 $^{1}A''$ (n→π*) | | | | | 5.12 | 5.07 | 5.02 | |

| | | | | |
|---|---|---|---|---|
| | 2 $^{1}A'$ (π→π*) | 5.54 | 5.64 | 5.58 |
| | 2 $^{1}A''$ (n→π*) | 5.66 | 5.59 | 5.53 |
| | 3 $^{1}A'$ (π→π*) | 6.43 | 6.50 | 6.51 |
| | 4 $^{1}A'$ (π→π*) | 6.86 | 6.91 | 6.85 |
| | 5 $^{1}A'$ (π→π*) | 7.73 | 7.71 | 7.68 |
| | 6 $^{1}A'$ (π→π*) | 8.18 | 8.16 | 8.12 |
| Thymine | 1 $^{1}A''$ (n→π*) | 4.87 | 4.75 | 4.67 |
| | 1 $^{1}A'$ (π→π*) | 5.13 | 5.24 | 5.24 |
| | 2 $^{1}A''$ (n→π*) | 6.28 | 6.19 | 6.14 |
| | 2 $^{1}A'$ (π→π*) | 6.37 | 6.44 | 6.36 |
| | 3 $^{1}A'$ (π→π*) | 6.67 | 6.76 | 6.71 |
| | 3 $^{1}A''$ (n→π*) | 6.70 | 6.67 | 6.63 |
| | 4 $^{1}A''$ (n→π*) | 7.25 | 7.25 | 7.21 |
| | 4 $^{1}A'$ (π→π*) | 7.40 | 7.51 | 7.50 |
| | 5 $^{1}A'$ (π→π*) | 8.33 | 8.33 | 8.30 |
| Uracil | 1 $^{1}A''$ (n→π*) | 4.87 | 4.74 | 4.66 |
| | 1 $^{1}A'$ (π→π*) | 5.18 | 5.33 | 5.34 |
| | 2 $^{1}A''$ (n→π*) | 6.19 | 6.11 | 6.08 |
| | 2 $^{1}A'$ (π→π*) | 6.37 | 6.43 | 6.34 |
| | 3 $^{1}A'$ (π→π*) | 6.82 | 6.91 | 6.86 |
| | 3 $^{1}A''$ (n→π*) | 6.87 | 6.84 | 6.80 |
| | 4 $^{1}A'$ (π→π*) | 7.26 | 7.43 | 7.44 |
| | 4 $^{1}A''$ (n→π*) | 7.26 | 7.27 | 7.22 |
| | 5 $^{1}A'$ (π→π*) | 8.29 | 8.28 | 8.26 |
| Adenine | 1 $^{1}A'$ (π→π*) | 4.87 | 5.02 | 4.98 |
| | 1 $^{1}A''$ (n→π*) | 5.24 | 5.18 | 5.15 |
| | 2 $^{1}A'$ (π→π*) | 5.39 | 5.36 | 5.29 |
| | 2 $^{1}A''$ (n→π*) | 5.86 | 5.78 | 5.73 |
| | 3 $^{1}A'$ (π→π*) | 6.56 | 6.55 | 6.49 |
| | 4 $^{1}A'$ (π→π*) | 6.68 | 6.77 | 6.68 |
| | 5 $^{1}A'$ (π→π*) | 6.85 | 6.84 | 6.73 |
| | 6 $^{1}A'$ (π→π*) | 7.42 | 7.40 | 7.31 |
| | 7 $^{1}A'$ (π→π*) | 7.81 | 7.78 | 7.68 |

Table 2b: Vertical triplet excitation energies in eV for all statistically evaluated molecules.

| Molecule | State | CC3 (%T1) | NEVPT2 | CASPT2 | STEOM-CC | EOM-CCSD (%T1) |
|---|---|---|---|---|---|---|
| Ethene | 1 $^3B_{1u}$ (π→π*) | 4.48 (99.30) | 4.60 | 4.60 | 4.42 | 4.42 (99.4) |
| E-Butadiene | 1 $^3B_u$ (π→π*) | 3.32 (98.50) | 3.39 | 3.44 | 3.13 | 3.25 (98.9) |
| | 1 $^3A_g$ (π→π*) | 5.17 (98.90) | 5.28 | 5.16 | 4.99 | 5.15 (99.1) |
| E-Hexatriene | 1 $^3B_u$ (π→π*) | 2.69 (98.00) | 2.74 | 2.71 | 2.46 | 2.62 (98.6) |
| | 1 $^3A_g$ (π→π*) | 4.32 (98.40) | 4.40 | 4.31 | 4.18 | 4.28 (98.9) |
| E-Octatetraene | 1 $^3B_u$ (π→π*) | 2.30 (97.60) | 2.33 | 2.33 | 2.06 | 2.23 (98.5) |
| | 1 $^3A_g$ (π→π*) | 3.67 (98.10) | 3.73 | 3.69 | 3.51 | 3.62 (98.7) |
| Cyclopropene | 1 $^3B_2$ (π→π*) | 4.34 (99.10) | 4.56 | 4.35 | 4.08 | 4.30 (99.2) |
| | 1 $^3B_1$ (σ→π*) | 6.62 (98.10) | 6.58 | 6.51 | 6.55 | 6.66 (98.5) |
| Cyclopentadiene | 1 $^3B_2$ (π→π*) | 3.25 (98.50) | 3.33 | 3.28 | 3.07 | 3.18 (98.9) |
| | 1 $^3A_1$ (π→π*) | 5.09 (98.70) | 5.23 | 5.10 | 5.02 | 5.07 (99) |
| Norbornadiene | 1 $^3A_2$ (π→π*) | 3.72 (98.70) | 3.81 | 3.75 | 3.49 | 3.67 (99) |
| | 1 $^3B_2$ (π→π*) | 4.16 (99.00) | 4.31 | 4.22 | 3.89 | 4.09 (99.2) |
| Benzene | 1 $^3B_{1u}$ (π→π*) | 4.12 (98.70) | 4.33 | 4.17 | 3.52 | 3.94 (99) |
| | 1 $^3E_{1u}$ (π→π*) | 4.90 (97.00) | 5.00 | 4.90 | 4.80 | 4.97 (97.9) |
| | 1 $^3B_{2u}$ (π→π*) | 6.04 (98.20) | 5.54 | 5.76 | 5.97 | 6.00 (98.6) |
| | 1 $^3E_{2g}$ (π→π*) | 7.49 (94.90) | 7.61 | 7.41 | 7.53 | 7.73 (97.6) |
| Naphthalene | 1 $^3B_{2u}$ (π→π*) | 3.11 (97.30) | 3.28 | 3.20 | 2.71 | 2.99 (98.2) |
| | 1 $^3B_{3u}$ (π→π*) | 4.18 (93.20) | 4.27 | 4.29 | 4.13 | 4.27 (97.9) |
| | 1 $^3B_{1g}$ (π→π*) | 4.47 (96.90) | 4.59 | 4.55 | 4.22 | 4.44 (97.4) |
| | 2 $^3B_{2u}$ (π→π*) | 4.64 (97.80) | 4.73 | 4.71 | 4.46 | 4.67 (98.6) |
| | 2 $^3B_{3u}$ (π→π*) | 5.11 (96.80) | 4.53 | 5.00 | 5.03 | 5.10 (97.8) |
| | 1 $^3A_g$ (π→π*) | 5.52 (96.50) | 5.62 | 5.57 | 5.42 | 5.57 (97.7) |
| | 2 $^3B_{1g}$ (π→π*) | 6.48 (97.60) | 5.95 | 6.25 | 6.64 | 6.79 (98.3) |
| | 2 $^3A_g$ (π→π*) | 6.47 (97.90) | 6.25 | 6.42 | 6.66 | 6.81 (98.5) |
| | 3 $^3A_g$ (π→π*) | 6.79 (95.00) | 6.56 | 6.63 | 6.77 | 6.96 (97.3) |
| | 3 $^3B_{1g}$ (π→π*) | 6.76 (94.00) | 6.83 | 6.67 | 6.83 | 7.04 (97.3) |
| Furan | 1 $^3B_2$ (π→π*) | 4.17 (98.50) | 4.36 | 4.17 | 3.84 | 4.10 (98.9) |
| | 1 $^3A_1$ (π→π*) | 5.48 (98.20) | 5.67 | 5.49 | 5.28 | 5.48 (98.7) |
| Pyrrole | 1 $^3B_2$ (π→π*) | 4.48 (98.40) | 4.74 | 4.52 | 4.18 | 4.41 (98.8) |
| | 1 $^3A_1$ (π→π*) | 5.51 (97.80) | 5.70 | 5.53 | 5.46 | 5.54 (98.4) |
| Imidazole | 1 $^3A'$ (π→π*) | 4.69 (98.40) | 4.80 | 4.65 | 4.40 | 4.62 (98.8) |
| | 2 $^3A'$ (π→π*) | 5.79 (97.90) | 5.93 | 5.74 | 5.68 | 5.83 (98.5) |
| | 1 $^3A''$ (n→π*) | 6.37 (97.40) | 6.49 | 6.36 | 6.23 | 6.43 (98.3) |
| | 3 $^3A'$ (π→π*) | 6.55 (97.90) | 6.67 | 6.44 | 6.40 | 6.56 (98.4) |
| | $4\,^{3}A'$ ($\pi\rightarrow\pi^*$) | 7.42 (97.10) | 7.15 | 7.43 | 7.32 | 7.54 (98) |
| | $2\,^{3}A''$ ($n\rightarrow\pi^*$) | 7.51 (96.00) | 7.61 | 7.51 | 7.56 | 7.76 (97.6) |
| Pyridine | $1\,^{3}A_1$ ($\pi\rightarrow\pi^*$) | 4.25 (98.60) | 4.48 | 4.27 | 3.69 | 4.07 (99) |
| | $1\,^{3}B_1$ ($n\rightarrow\pi^*$) | 4.50 (97.10) | 4.60 | 4.55 | 4.39 | 4.61 (98.1) |
| | $1\,^{3}B_2$ ($\pi\rightarrow\pi^*$) | 4.86 (97.20) | 4.97 | 4.72 | 4.82 | 4.91 (98) |
| | $2\,^{3}A_1$ ($\pi\rightarrow\pi^*$) | 5.05 (97.00) | 5.16 | 5.03 | 5.00 | 5.13 (97.9) |
| | $1\,^{3}A_2$ ($n\rightarrow\pi^*$) | 5.46 (95.80) | 5.49 | 5.48 | 5.43 | 5.67 (97.5) |
| | $2\,^{3}B_2$ ($\pi\rightarrow\pi^*$) | 6.40 (97.80) | 6.49 | 6.02 | 6.34 | 6.41 (98.3) |
| | $3\,^{3}A_1$ ($\pi\rightarrow\pi^*$) | 7.66 (95.30) | 7.87 | 7.56 | 7.73 | 7.90 (97.7) |
| | $3\,^{3}B_2$ ($\pi\rightarrow\pi^*$) | 7.83 (94.40) | 7.28 | 7.88 | 7.94 | 8.12 (97.4) |
| Tetrazine | $1\,^{3}B_{3u}$ ($n\rightarrow\pi^*$) | 1.89 (97.20) | 1.69 | 1.56 | 1.79 | 1.99 (98.1) |
| | $1\,^{3}A_u$ ($n\rightarrow\pi^*$) | 3.52 (96.30) | 3.46 | 3.26 | 3.55 | 3.74 (97.7) |
| | $1\,^{3}B_{1u}$ ($\pi\rightarrow\pi^*$) | 4.33 (98.50) | 4.56 | 4.36 | 3.67 | 4.05 (99) |
| | $1\,^{3}B_{1g}$ ($n\rightarrow\pi^*$) | 4.21 (97.10) | 4.37 | 4.14 | 4.04 | 4.31 (98.2) |
| | $1\,^{3}B_{2u}$ ($\pi\rightarrow\pi^*$) | 4.54 (97.40) | 4.76 | 4.56 | 4.53 | 4.57 (98.1) |
| | $1\,^{3}B_{2g}$ ($n\rightarrow\pi^*$) | 4.93 (96.40) | 5.22 | 4.93 | 4.87 | 5.09 (98) |
| | $2\,^{3}A_u$ ($n\rightarrow\pi^*$) | 5.03 (96.60) | 5.11 | 5.02 | 5.01 | 5.20 (97.8) |
| | $2\,^{3}B_{1u}$ ($\pi\rightarrow\pi^*$) | 5.38 (96.50) | 5.54 | 5.40 | 5.42 | 5.48 (97.5) |
| | $2\,^{3}B_{2g}$ ($n\rightarrow\pi^*$) | 6.04 (93.00) | 6.18 | 5.97 | 6.29 | 6.51 (96.8) |
| | $2\,^{3}B_{3u}$ ($n\rightarrow\pi^*$) | 6.53 (95.80) | 6.78 | 6.54 | 6.63 | 6.80 (97.5) |
| | $2\,^{3}B_{1g}$ ($n\rightarrow\pi^*$) | 6.60 (92.30) | 6.62 | 6.31 | 6.86 | 7.11 (96.9) |
| | $2\,^{3}B_{2u}$ ($\pi\rightarrow\pi^*$) | 7.36 (96.80) | 6.54 | 7.10 | 7.52 | 7.46 (97.7) |
| Formaldehyde | $1\,^{3}A_2$ ($n\rightarrow\pi^*$) | 3.55 (98.10) | 3.75 | 3.58 | 3.50 | 3.52 (98.6) |
| | $1\,^{3}A_1$ ($\pi\rightarrow\pi^*$) | 5.83 (99.20) | 6.05 | 5.84 | 5.65 | 5.78 (99.3) |
| Acetone | $1\,^{3}A_2$ ($n\rightarrow\pi^*$) | 4.05 (97.90) | 4.13 | 4.08 | 3.95 | 4.03 (98.4) |
| | $1\,^{3}A_1$ ($\pi\rightarrow\pi^*$) | 6.03 (98.90) | 6.04 | 6.03 | 5.81 | 5.94 (99.1) |
| Benzoquinone | $1\,^{3}B_{1g}$ ($n\rightarrow\pi^*$) | 2.51 (95.90) | 2.88 | 2.63 | 2.54 | 2.71 (97.9) |
| | $1\,^{3}B_{1u}$ ($\pi\rightarrow\pi^*$) | 2.96 (97.80) | 2.94 | 2.99 | 2.64 | 2.89 (98.5) |
| | $1\,^{3}A_u$ ($n\rightarrow\pi^*$) | 2.62 (95.70) | 2.89 | 2.68 | 2.65 | 2.83 (97.8) |
| | $1\,^{3}B_{3g}$ ($\pi\rightarrow\pi^*$) | 3.41 (98.00) | 3.43 | 3.31 | 3.20 | 3.42 (98.6) |
| Formamide | $1\,^{3}A''$ ($n\rightarrow\pi^*$) | 5.36 (97.80) | 5.65 | 5.40 | 5.23 | 5.32 (98.4) |
| | $1\,^{3}A'$ ($\pi\rightarrow\pi^*$) | 5.74 (98.40) | 5.87 | 5.58 | 5.63 | 5.67 (98.7) |
| Acetamide | $1\,^{3}A''$ ($n\rightarrow\pi^*$) | 5.42 (98.30) | 5.54 | 5.53 | 5.26 | 5.39 (98.4) |
| | $1\,^{3}A'$ ($\pi\rightarrow\pi^*$) | 5.88 (98.30) | 5.67 | 5.75 | 5.73 | 5.83 (98.7) |
| Propanamide | $1\,^{3}A''$ ($n\rightarrow\pi^*$) | 5.45 (97.70) | 5.60 | 5.44 | 5.27 | 5.41 (98.4) |
| | $1\,^{3}A'$ ($\pi\rightarrow\pi^*$) | 5.90 (98.30) | 5.90 | 5.79 | 5.74 | 5.84 (98.7) |

In the following sections, we analyze our benchmark results and draw conclusions regarding STEOM methods.

### III.B.1. CC3 against EOM-CCSDT-3

In this subsection we analyze the benchmark results for CC3 and EOM-CCSDT-3 methods for the singlet excitations. For the triplet excitations, as reported by Schreiber et al., EOM-CCSD results are very close to CC3 results. This reflects that electron correlation effects are less pronounced for triplet excited states. For this reason one can expect also EOM-CCSDT-3 results to be very close for the triplet states, and no further analysis was attempted. A critical parameter that is used to gauge the reliability of CC3 calculations is the %T1 diagnostic. If this parameter becomes too small, double excitations are important, and correspondingly the perturbative treatment of triples corrections in CC3 or EOM-CCSDT-3 becomes suspect. We use a %T1 threshold value 87%, meaning that only states that have %T1 larger than or equal to this value are included in the statistical analysis. All excitation energies are reported in Tables 2a and 2b regardless of %T1. Our %T1 threshold is somewhat conservative probably, and 85% has been used also in the literature [72].

We present our statistical data for the comparison between CC3 and EOM-CCSDT-3 in Table 3.

| **Statistics** | **EOM-CCSDT-3 − CC3** |
|---|---|
| Minimum Error | 0.006 |
| Maximum Error | 0.113 |
| Mean Error | 0.049 |
| Mean Absolute Error | 0.049 |
| Root Mean Square Error | 0.056 |
| Standard Deviation | 0.027 |

Table 3: Statistical analysis of the benchmark set excitation energy deviations in eV of EOM-CCSDT-3 from CC3 for singlet states with T1≥87%.

We also perform a histogram bar chart analysis for our results (Figure 2a). In such a chart one counts the number of occurrences (errors or deviations here) in given equally spaced intervals. The bin size determines our distribution as the central interval is always centered around zero. This bin size is reported in the figure caption. To get a qualitative picture of the distribution, we also construct continuous plots. This is a more convenient representation if a number of distributions are examined in a single graph. We emphasize that continuous plots (using 3-point spline extrapolation) are not an entirely accurate representation of our results, as our results are constituted of discrete data points. In all following subsections, we use continuous plots to represent our data as they provide significant ease and insight of the distribution studied. We also note that the distribution is sensitive to the bin size chosen. We always attempt to choose a convenient bin size that will provide a suitable representation of the data.

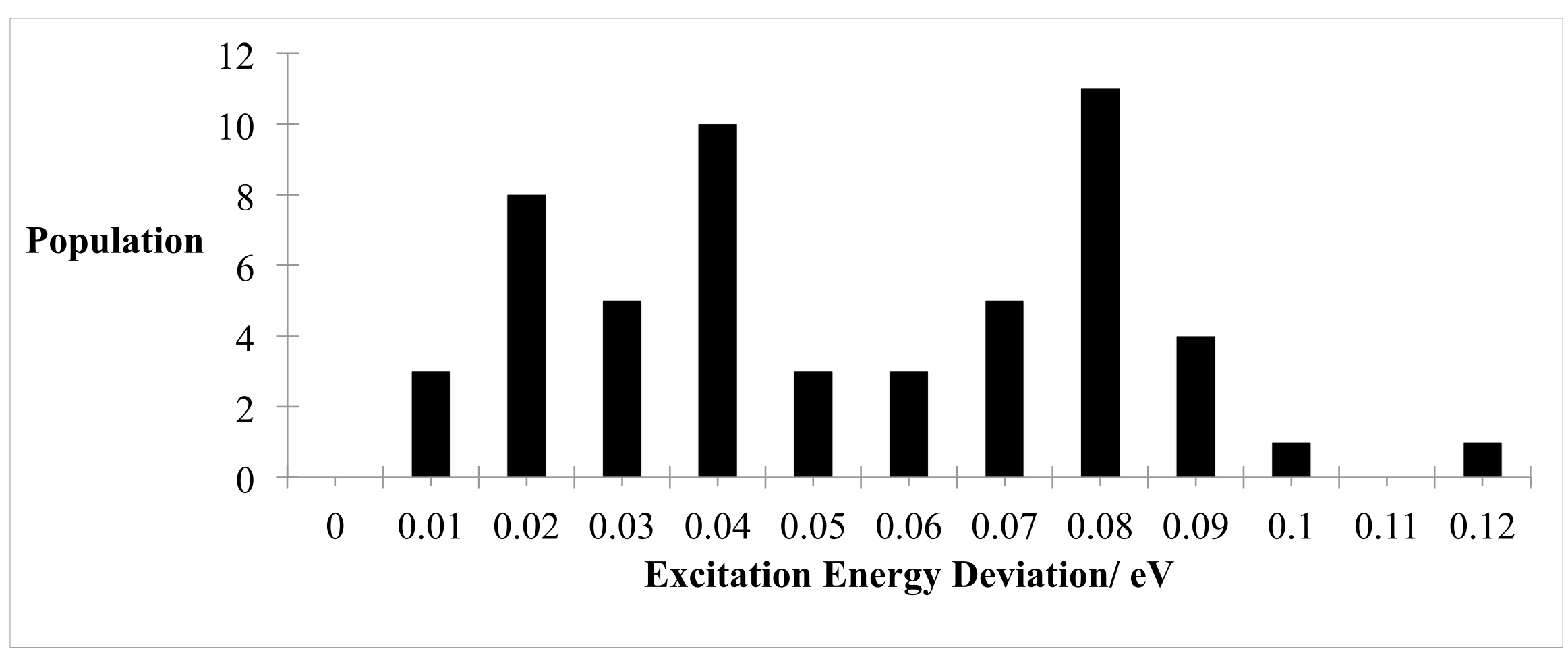

Figure 2a: Histogram bar chart (bin size 0.01), showing distribution of the benchmark set excitation energy deviations of EOM-CCSDT-3 from CC3 for singlet states with %T1≥87%.

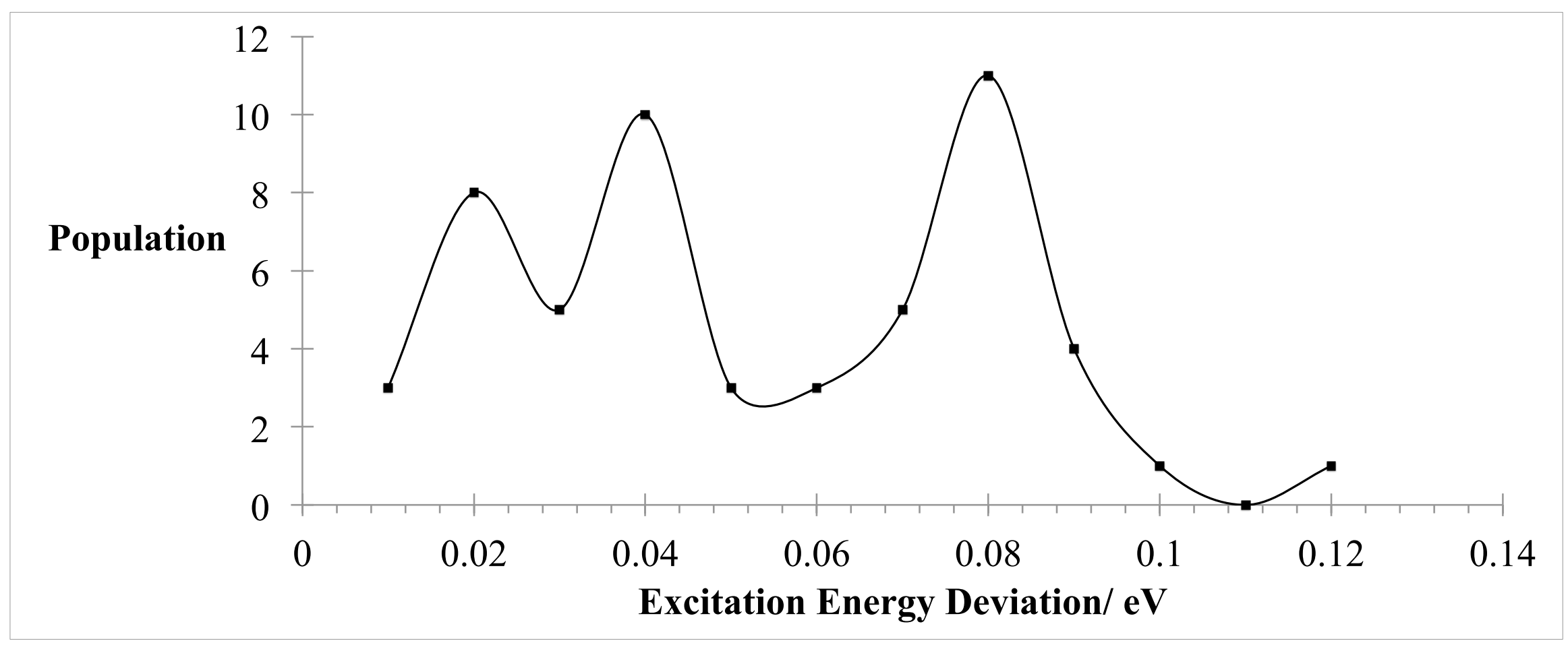

Figure 2b: Continuous curves as a representation for the distribution of the benchmark set excitation energy deviations of EOM-CCSDT-3 from CC3 for singlet states with T1≥87%.

The results presented here are somewhat surprising. From previous CC3 benchmark studies on small molecules it has been established that CC3 results can be expected to be quite close to the full Configuration Interaction (CI) solution, within a mean absolute error of only 0.016 eV for singles dominated singlet as well as triplet excitation energies [105]. From a formal theoretical perspective the EOM-CCSDT-3 approach is slightly more complete than CC3, and is likewise expected to be close to the full CI. Nonetheless, for the present test set of larger molecules, we find somewhat significant deviations of around 0.08 eV, quite frequently. We note that all deviations greater than 0.07 eV are $\pi \rightarrow \pi^*$ excitations with the exception of two deep lying states for pyridazine and tetrazine, which are $n \rightarrow \pi^*$ excitations. Since these methods serve to benchmark our STEOM results, and other approaches in the literature, e.g. [93, 96], these deviations are larger than we would like. These results are indicative that a future more accurate benchmark is desired. A possible candidate for such a future benchmark might be the CC($P$;$Q$) approach developed by Shen and Piecuch [39, 40] or the full EOM-CCSDT approach [106, 107].

### III.B.2. STEOM-H (ω) Methods

As discussed in section II., a hermitized version of STEOM is of interest, and we discussed a continuous family of approaches denoted STEOM-H (ω), with the $\omega \rightarrow \infty$ limit representing the simplest averaging of the transformed Hamiltonian and its transpose. Here we present the results of the statistical analysis of the benchmarks of different STEOM-H (ω) methods compared to STEOM-CC in Tables 4a and 4b for singlet excitations and triplet excitations respectively.

| **Statistics** | **STEOM-H (ω*)** | **STEOM-H (0)** | **STEOM-H (0.5)** | **STEOM-H (∞)** |
|---|---|---|---|---|
| Minimum Error | -0.103 | -0.073 | -0.098 | -0.124 |
| Maximum Error | 0.000 | -0.003 | -0.004 | -0.012 |
| Mean Error | -0.016 | -0.023 | -0.033 | -0.048 |
| Mean Absolute Error | 0.016 | 0.023 | 0.033 | 0.048 |
| Root Mean Square Error | 0.020 | 0.025 | 0.037 | 0.052 |
| Standard Deviation | 0.012 | 0.012 | 0.016 | 0.020 |

Table 4a: Statistical analysis of the benchmark set excitation energy deviations in eV of STEOM-H (ω) methods from STEOM-CC for singlet states.

| Statistics | STEOM-H (ω*) | STEOM-H (0) | STEOM-H (0.5) | STEOM-H (∞) |
|---|---|---|---|---|
| Minimum Error | -0.026 | -0.036 | -0.053 | -0.071 |
| Maximum Error | 0.004 | -0.002 | -0.005 | -0.010 |
| Mean Error | -0.006 | -0.013 | -0.023 | -0.037 |
| Mean Absolute Error | 0.007 | 0.013 | 0.023 | 0.037 |
| Root Mean Square Error | 0.009 | 0.014 | 0.025 | 0.039 |
| Standard Deviation | 0.006 | 0.006 | 0.010 | 0.014 |

Table 4b: Statistical analysis of the benchmark set excitation energy deviations in eV of STEOM-H (ω) methods from STEOM-CC for triplet states.

As in the previous section, we perform a histogram analysis for the results and we represent the distributions as continuous curves (Figures 3a and 3b).

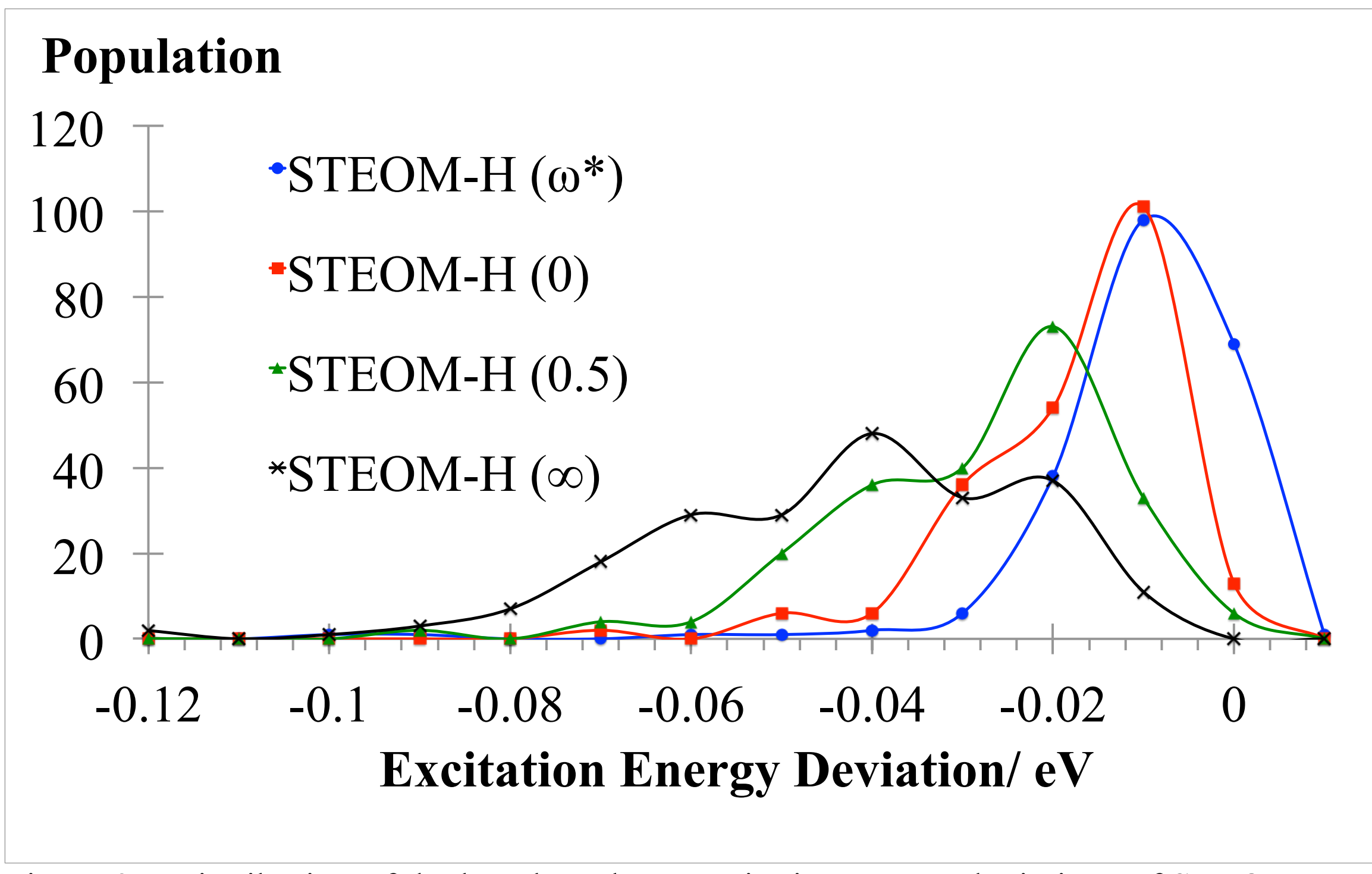

Figure 3a: Distribution of the benchmark set excitation energy deviations of STEOM-H (ω) methods from STEOM-CC for singlet states. Bin size: 0.01.

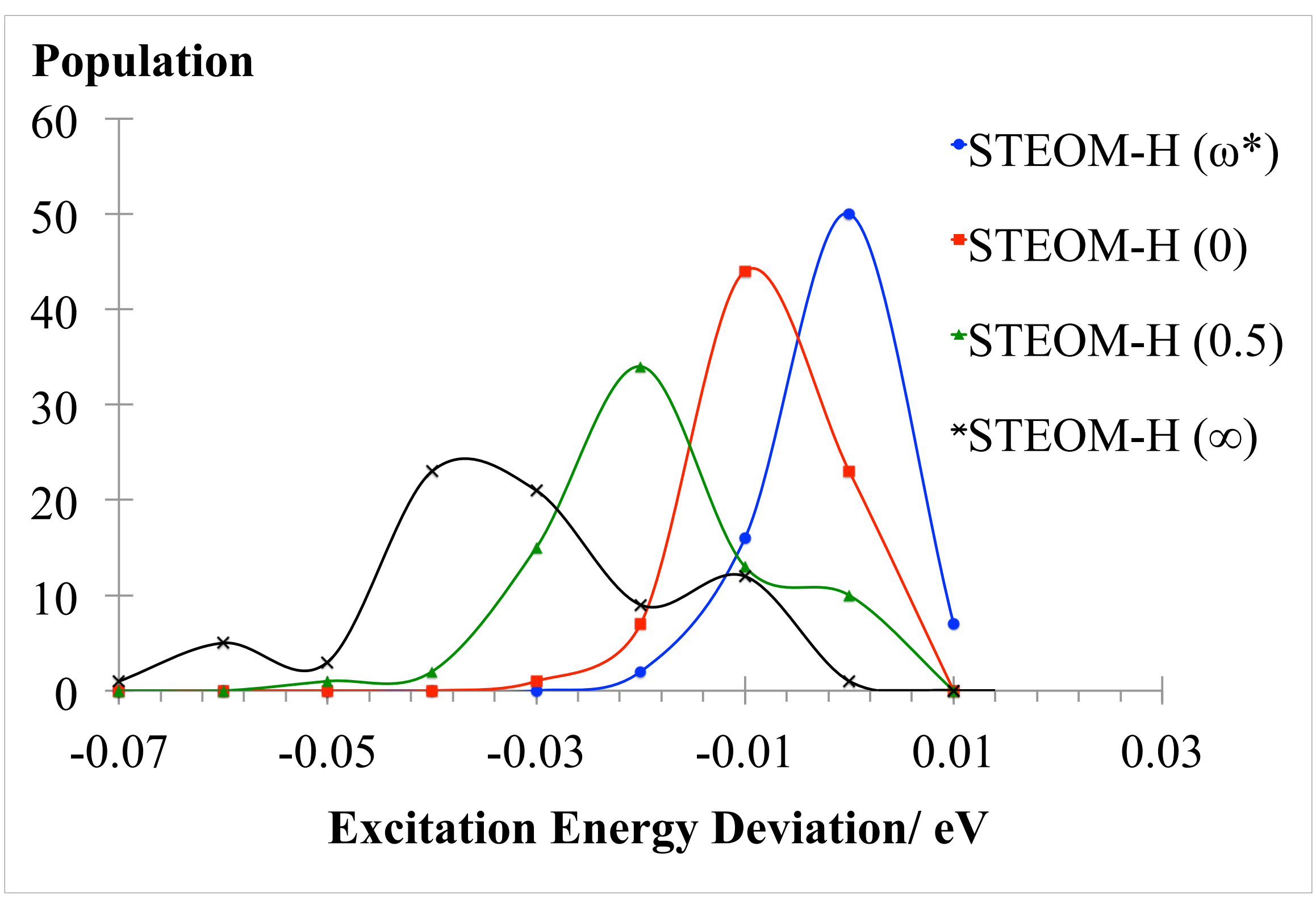

Figure 3b: Distribution of the benchmark set excitation energy deviations of STEOM-H (ω) methods from STEOM-CC for triplet states. Bin size: 0.01.

The best STEOM-H method is unambiguously the STEOM-H (ω*) approach. Let us recall that in this approach the ω parameter is determined for each symmetry block of excitation energies separately from the diagonal elements of the sector of the transformed Hamiltonian. In practice the value of ω* is between -0.1 a.u. or -2.7 eV and -0.2 a.u. or -5.4 eV. The deviation from STEOM-CC monotonically increases as ω increases with greatest deviation for $\omega \rightarrow \infty$ as is evident from Figures 3a, 3b. Since these results are quite close to STEOM-CC in general, no further analysis of the hermitized approaches is needed. Their trends will clearly follow the parent STEOM-CC approach.

### III.B.3. STEOM-ORB

Next we investigate another minor issue: the dependence of STEOM-CC results on the precise choice of active orbitals. In the STEOM-ORB approach a set of "natural orbitals" is introduced within the subspaces of occupied and virtual Hartree-Fock orbitals, as discussed in section II. We present the results of the statistical analysis of the benchmarks of STEOM-ORB compared to STEOM-CC in Table 5 for singlet excitations and triplet excitations.

| Statistics | Singlets | Triplets |
|---|---|---|
| Minimum Error | -0.055 | -0.043 |
| Maximum Error | 0.004 | 0.002 |
| Mean Error | -0.008 | -0.014 |
| Mean Absolute Error | 0.008 | 0.014 |
| Root Mean Square Error | 0.011 | 0.018 |
| Standard Deviation | 0.008 | 0.011 |

Table 5: Statistical analysis of the benchmark set excitation energy deviations in eV of STEOM-ORB from STEOM-CC for singlet states and triplet states.

The distributions are represented as continuous curves in Figures 4a and 4b below.

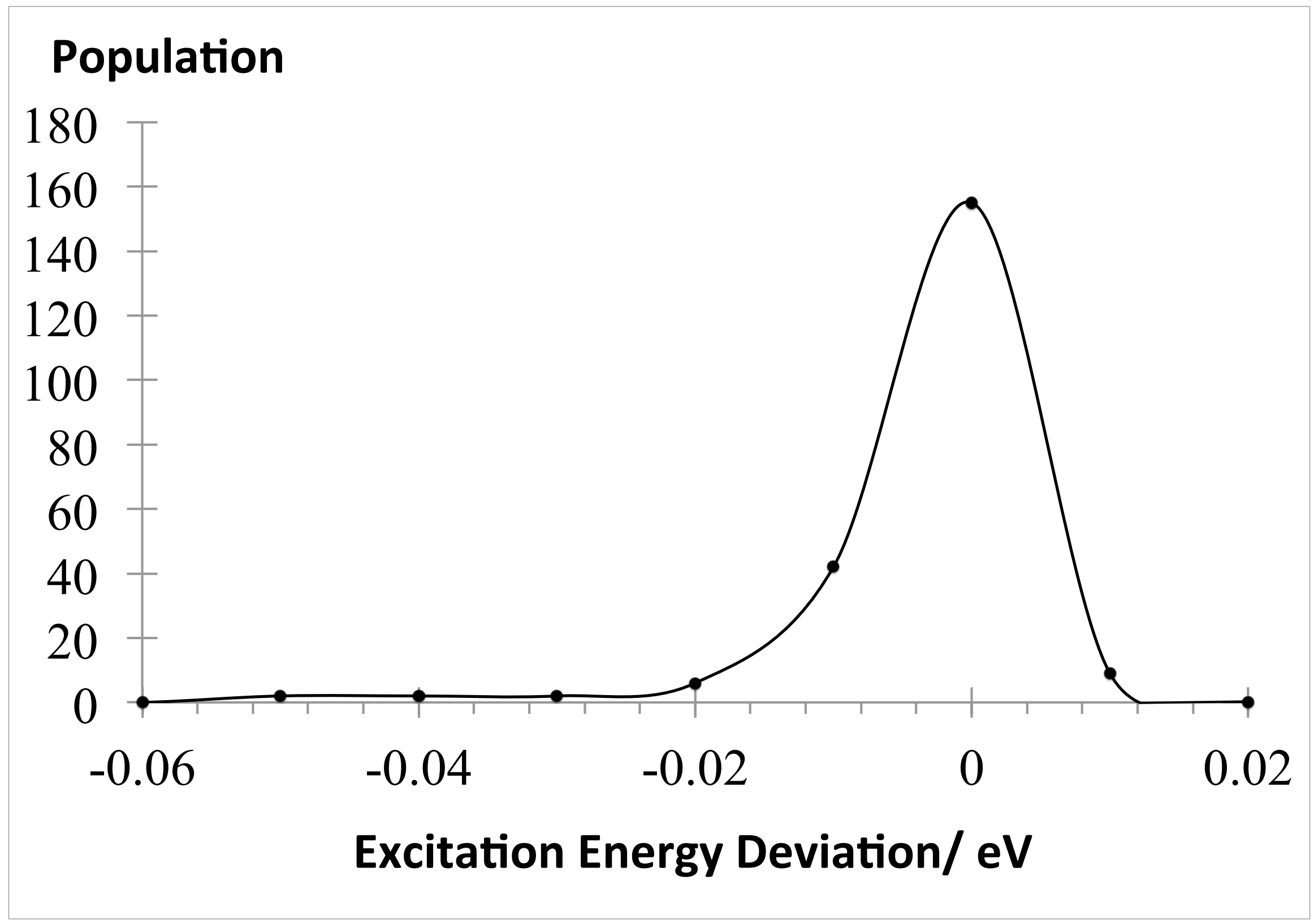

Figure 4a: Distribution of the benchmark set excitation energy deviations of STEOM-ORB from STEOM-CC for singlet states. Bin size: 0.01.

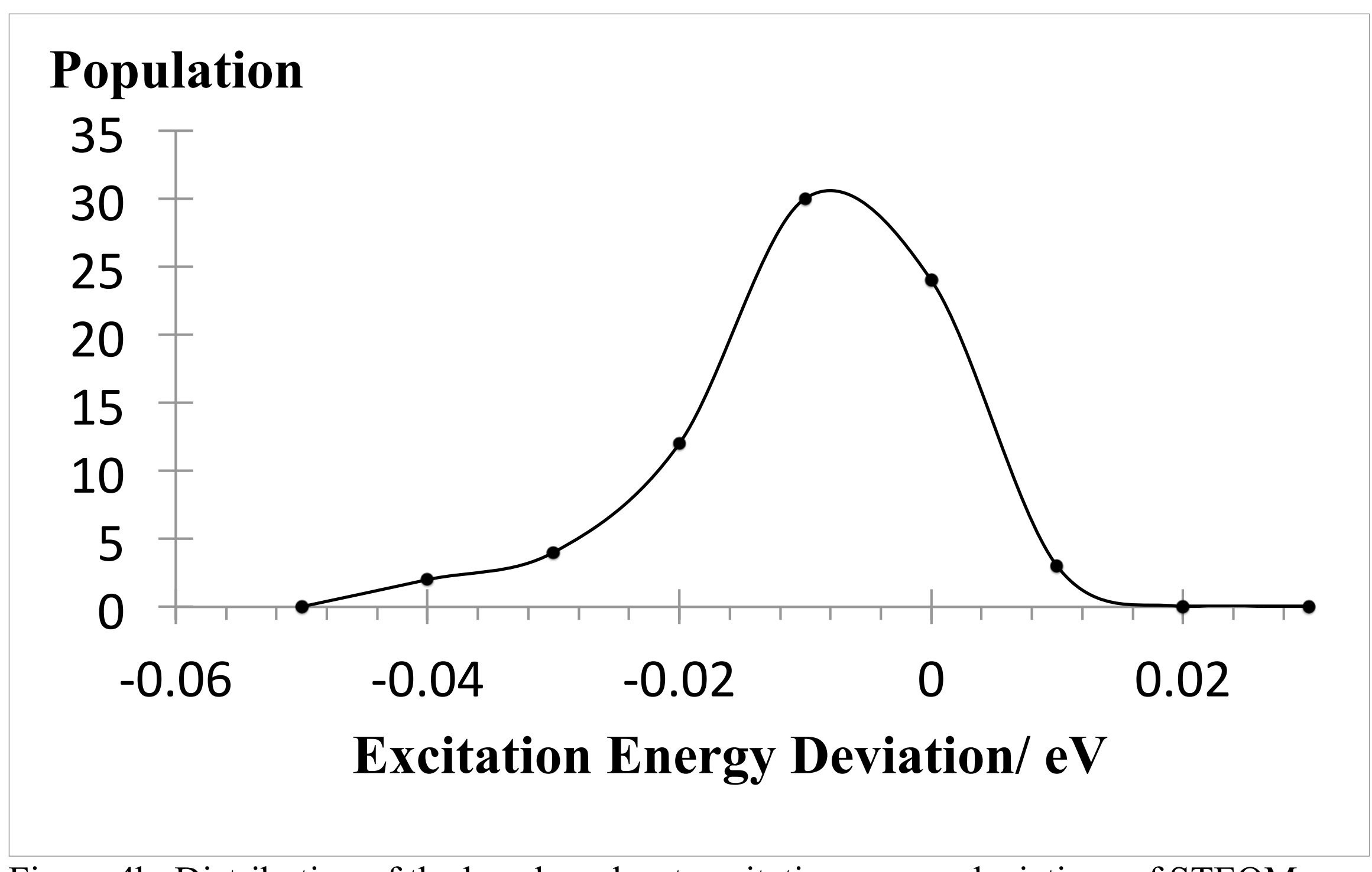

Figure 4b: Distribution of the benchmark set excitation energy deviations of STEOM-ORB from STEOM-CC for triplet states. Bin size: 0.01.

These results clearly indicate that STEOM-CC results are quite insensitive to the precise definition of active orbitals. This is a redeeming feature of the approach. Results are clear-cut again, and there is no need to make additional analysis of the STEOM-ORB approach. They closely follow the parent STEOM-CC results.

### III.B.4. STEOM-PT

We study the results of the statistical analysis of the benchmarks of STEOM-PT compared to STEOM-CC in Table 6 for singlet excitations and triplet excitations.

| Statistics | Singlets | Triplets |
|---|---|---|
| Minimum Error | -0.265 | -0.188 |
| Maximum Error | 0.264 | 0.442 |
| Mean Error | 0.056 | 0.064 |
| Mean Absolute Error | 0.099 | 0.093 |
| Root Mean Square Error | 0.121 | 0.120 |
| Standard Deviation | 0.107 | 0.102 |

Table 6: Statistical analysis of the benchmark set excitation energy deviations in eV of STEOM-PT from STEOM-CC for singlet states and triplet states.

We demonstrate the distributions as continuous curves (Figures 5a and 5b) to provide insight about the STEOM-PT methodology.

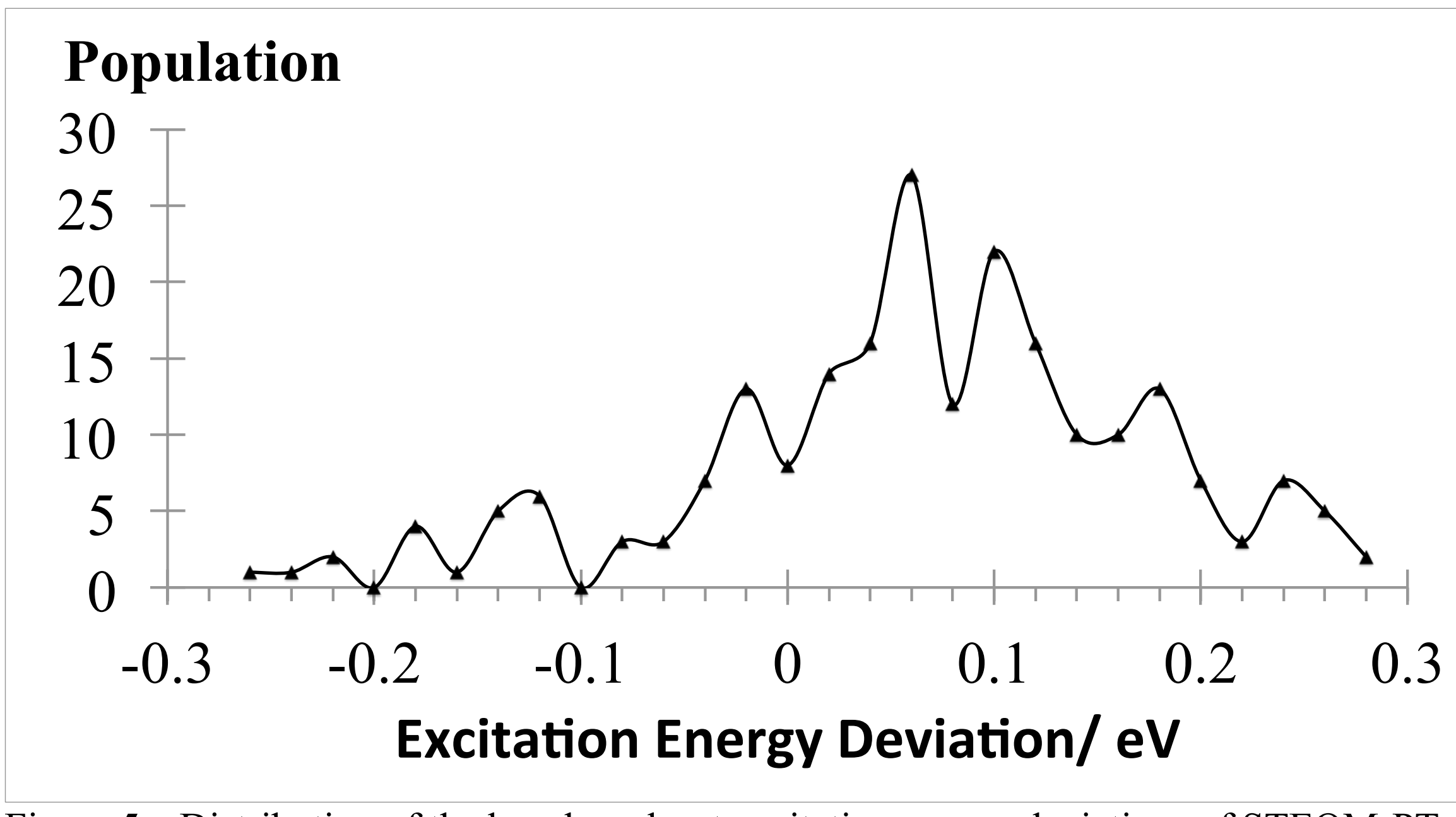

Figure 5a: Distribution of the benchmark set excitation energy deviations of STEOM-PT from STEOM-CC for singlet states. Bin size: 0.02.

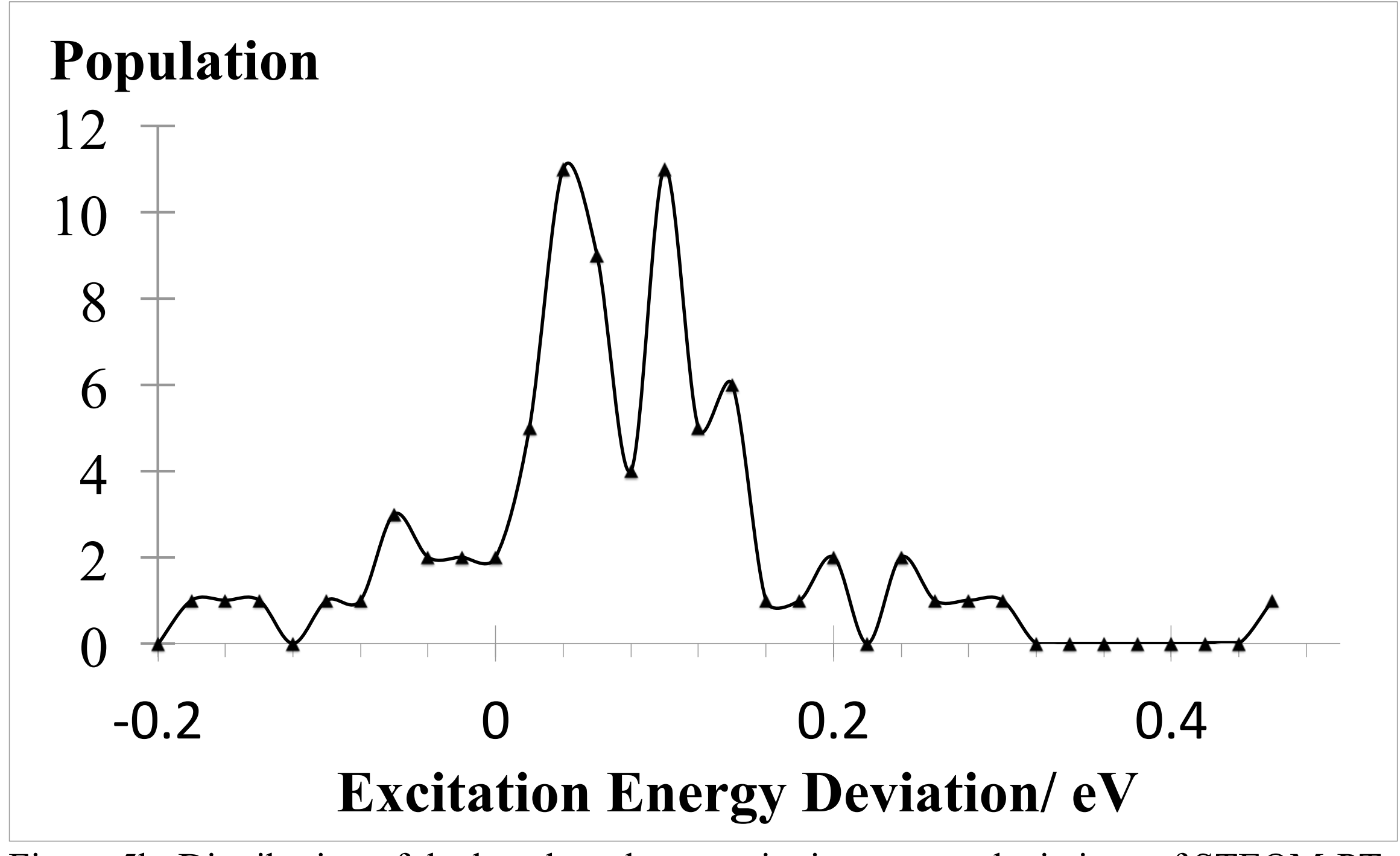

Figure 5b: Distribution of the benchmark set excitation energy deviations of STEOM-PT from STEOM-CC for triplet states. Bin size: 0.02.

The deviations of STEOM-PT from STEOM-CC for singlet states are rather erratic and the distribution for deviations is quite spread out with some sizeable deviations. For triplet states, the distribution of deviations is likewise rather broad. Given these results, and the fact that STEOM-CC is only about twice as demanding as STEOM-PT, the STEOM-PT approach is not all that interesting. Only if the calculation of the EA amplitudes would be approximated with a cheaper method like for example with the partitioned EOM-CC approach [108], results that might have merits should prove interesting.

### III.B.5. STEOM-CC, STEOM-D and EOM-CCSD against CC3 and EOM-CCSDT-3

In this section we discuss the accuracy of STEOM-CC, STEOM-D and EOM-CCSD against CC3 and EOM-CCSDT-3. We commence by discussing results for singlet excitations and then we discuss results for triplet excitations.

#### III.B.5.1 Singlet Excitations

We present the results of the statistical analysis of the benchmarks of STEOM-CC, STEOM-D, EXT-STEOM and EOM-CCSD compared to CC3 and EOM-CCSDT-3 (separated by a slash) in Table 7 and we plot continuous distributions in Figures 6a and 6b. Again, we only include CC3 and EOM-CCSDT-3 results with T1≥87% for the reasons mentioned in section III.B.1.

| Statistics | STEOM-CC | STEOM-D | EXT-STEOM | EOM-CCSD |
|---|---|---|---|---|
| Minimum Error | -0.219 / -0.229 | -0.323 / -0.334 | -0.374 / -0.384 | 0.009 / -0.002 |
| Maximum Error | 0.159 / 0.120 | 0.100 / -0.010 | 0.083 / -0.055 | 0.408 / 0.255 |
| Mean Error | -0.019 / -0.073 | -0.067 / -0.120 | -0.118 / -0.161 | 0.198 /0.125 |
| Mean Absolute Error | 0.065 / 0.091 | 0.085 / 0.120 | 0.123 / 0.161 | 0.198 / 0.125 |
| Root Mean Square Error | 0.081 / 0.105 | 0.118 / 0.146 | 0.160 / 0.189 | 0.216 / 0.136 |
| Standard Deviation | 0.079 / 0.076 | 0.097 / 0.084 | 0.109 / 0.100 | 0.086 / 0.056 |

Table 7: Statistical analysis of the benchmark set excitation energy deviations in eV of STEOM-CC, STEOM-D, EXT-STEOM and EOM-CCSD from CC3 and from EOM-CCSDT-3 for singlet states. Only CC3 and EOM-CCSDT-3 states with T1≥87% are included in the analysis. A slash separates the values for the comparison against CC3 from that against EOM-CCSDT-3.

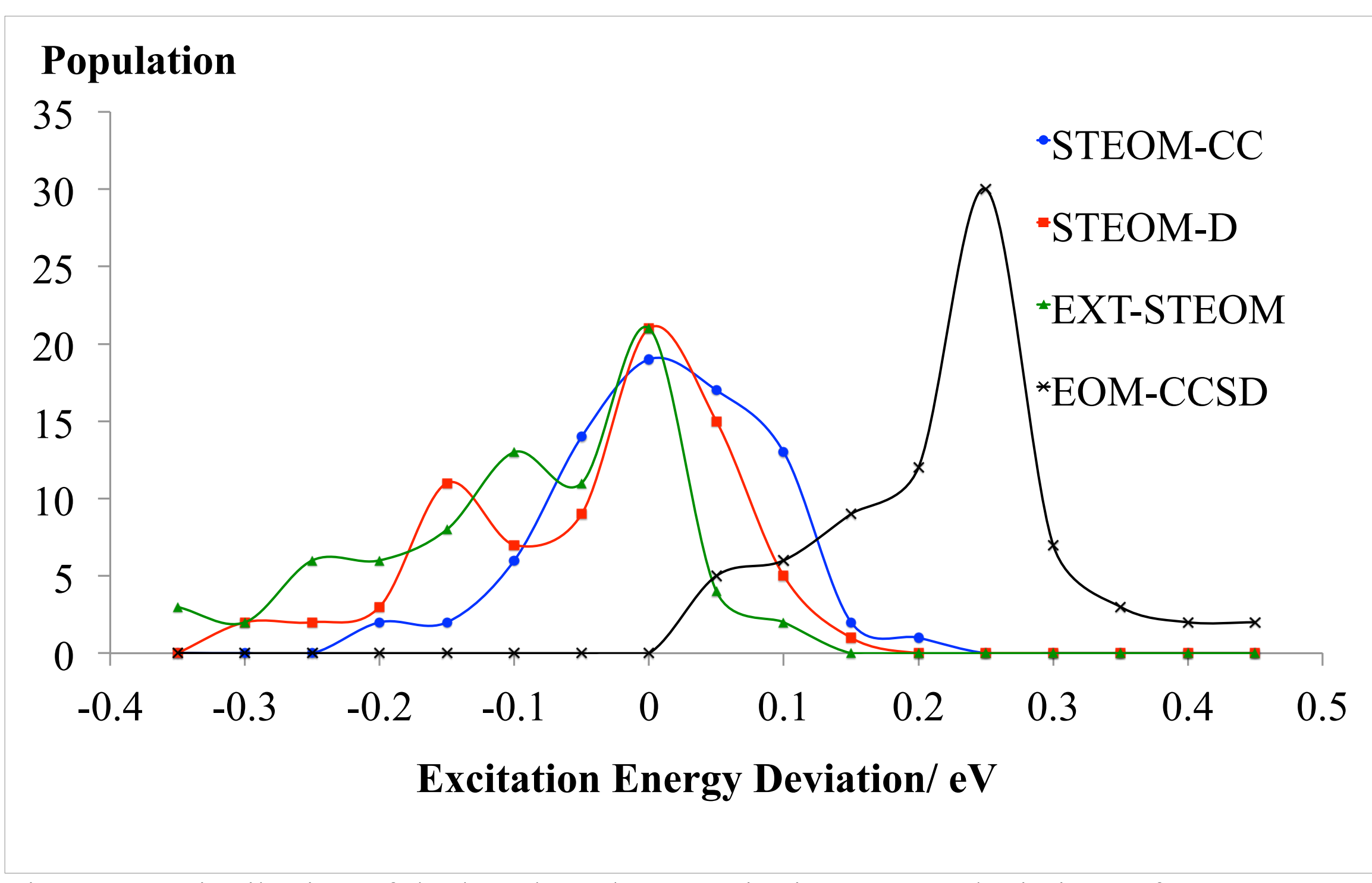

Figure 6a: Distribution of the benchmark set excitation energy deviations of STEOM-CC, STEOM-D, EXT-STEOM and EOM-CCSD from CC3 for singlet states. Only CC3 states with T1≥87% are included in the analysis. Bin size: 0.05.

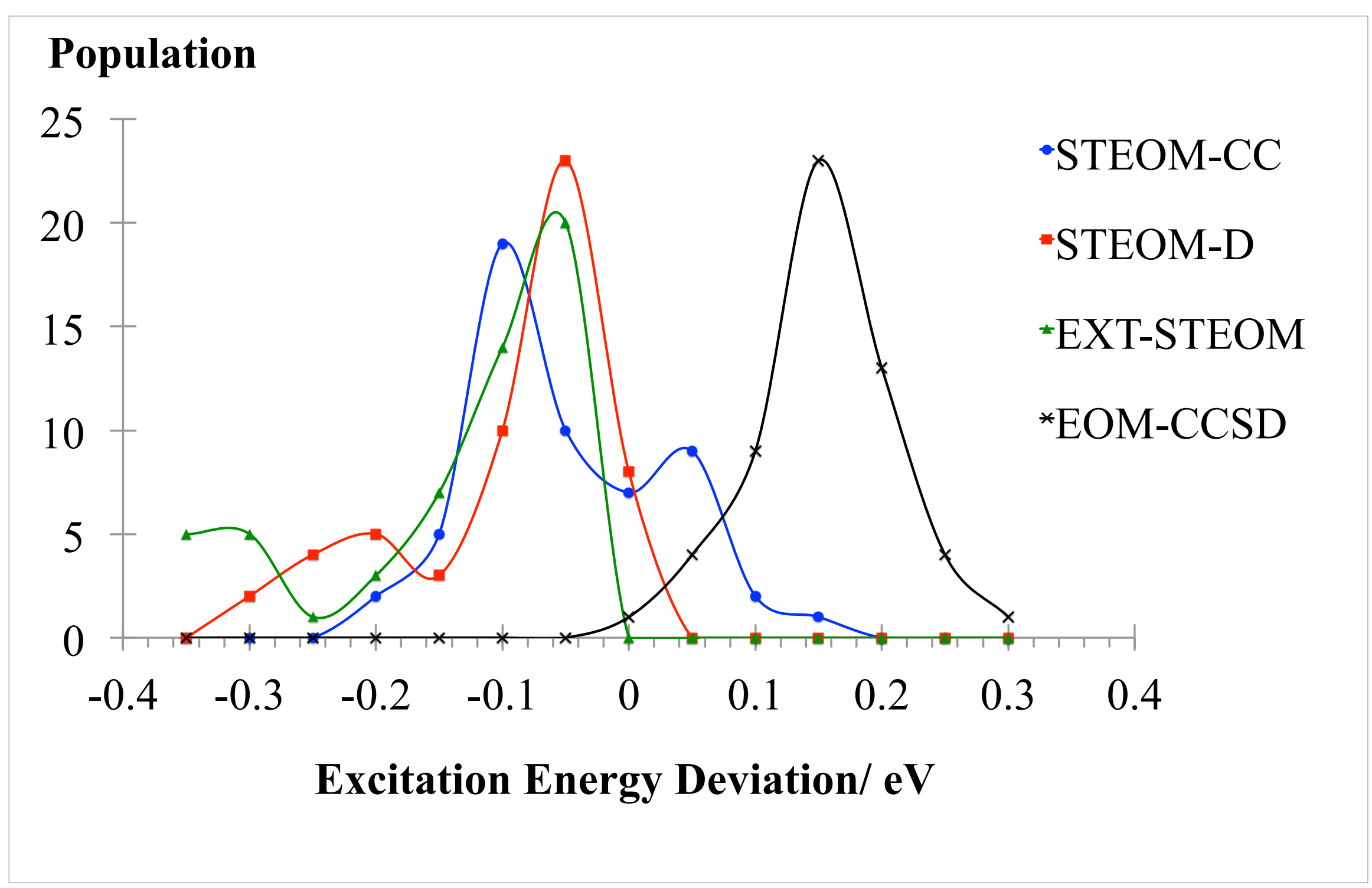

Figure 6b: Distribution of the benchmark set excitation energy deviations of STEOM-CC, STEOM-D, EXT-STEOM and EOM-CCSD from EOM-CCSDT-3 for singlet states. Only EOM-CCSDT-3 states with T1≥87% are included in the analysis. Bin size: 0.05.

The deviation of EOM-CCSD is very systematic in particular in comparison to EOM-CCSDT-3. This is clear from the histogram, which exhibits a sharply peaked distribution (standard deviation of 0.056 eV), centered at 0.15 eV (0.125 eV mean absolute error). The comparison with CC3 is somewhat less systematic.

For STEOM-CC the distribution is somewhat broader than for EOM-CCSD (standard deviation of 0.079 eV/ 0.076 eV). It is centered more closely to zero, in particular when compared to CC3. Overall the most relevant root mean square error is better for STEOM-CC than for EOM-CCSD.

The statistical analysis indicates that STEOM-CC is more accurate than STEOM-D. However, the histogram analysis as presented in the figures above provide additional information, specifically that STEOM-D is quite accurate in general with the exception of some outliers. It is interesting to see that the STEOM-D has a sharper peak than STEOM-CC, in particular when compared to EOM-CCSDT-3. The distribution is shifted to about 0.05 eV lower excitation energies compared to EOM-CCSDT-3 with a wing at a deviation of around -0.2 eV. This wing corresponds to outliers, which are mostly $n \rightarrow \pi^*$ excitations involving 'double bond O' groups found in amides, ketones and aldehydes. The only exceptions to these outliers are two $\sigma \rightarrow \pi^*$ excitations.

EXT-STEOM has a broader distribution than STEOM-CC and STEOM-D. The distribution has a pronounced tail with a number of states exhibiting deviations up to about -0.4 eV. The fact that STEOM-CC and EXT-STEOM yield quite comparable results, while the size of the diagonalization manifold is greatly different, testifies to the effectiveness of the similarity transform approach. It is natural that EXT-STEOM yields lower excitation energies, as the ground state CCSD energy is the same in both approaches, while the diagonalization manifold is much larger in EXT-STEOM.
Given the low computational cost of STEOM methods and the statistical results, we can say that the STEOM approach is behaving very satisfactorily when comparing to the computationally more expensive EOM-CCSD.

### III.B.5.2 Triplet Excitations

We present the results of the statistical analysis of the benchmarks of STEOM-CC, and EOM-CCSD compared to CC3 in Table 8 for triplet excitations and we present the distributions in Figure 7. First, we note that STEOM-D and EXT-STEOM methods are not implemented for triplet states, hence their absence in this section. We remind the reader that we do not benchmark EOM-CCSDT-3 for triplet states as we expect EOM-CCSDT-3 to be very close to CC3. The reader is referred to the discussion in section III.B.1. We also note that the CC3 results for our benchmark set has a relatively high %T1 being in general above 92%. The excitation energy results are found in Table 2b. For this reason, we do not employ the %T1 diagnostic to filter our results and we simply include all the results in the statistical analysis.

| **Statistics** | **STEOM-CC** | **EOM-CCSD** |
|---|---|---|
| Minimum Error | -0.662 | -0.280 |
| Maximum Error | 0.256 | 0.510 |
| Mean Error | -0.111 | 0.053 |
| Mean Absolute Error | 0.156 | 0.114 |
| Root Mean Square Error | 0.201 | 0.156 |
| Standard Deviation | 0.169 | 0.148 |

Table 8: Statistical analysis of the benchmark set excitation energy deviations in eV of STEOM-CC and EOM-CCSD from CC3 for Triplet states.

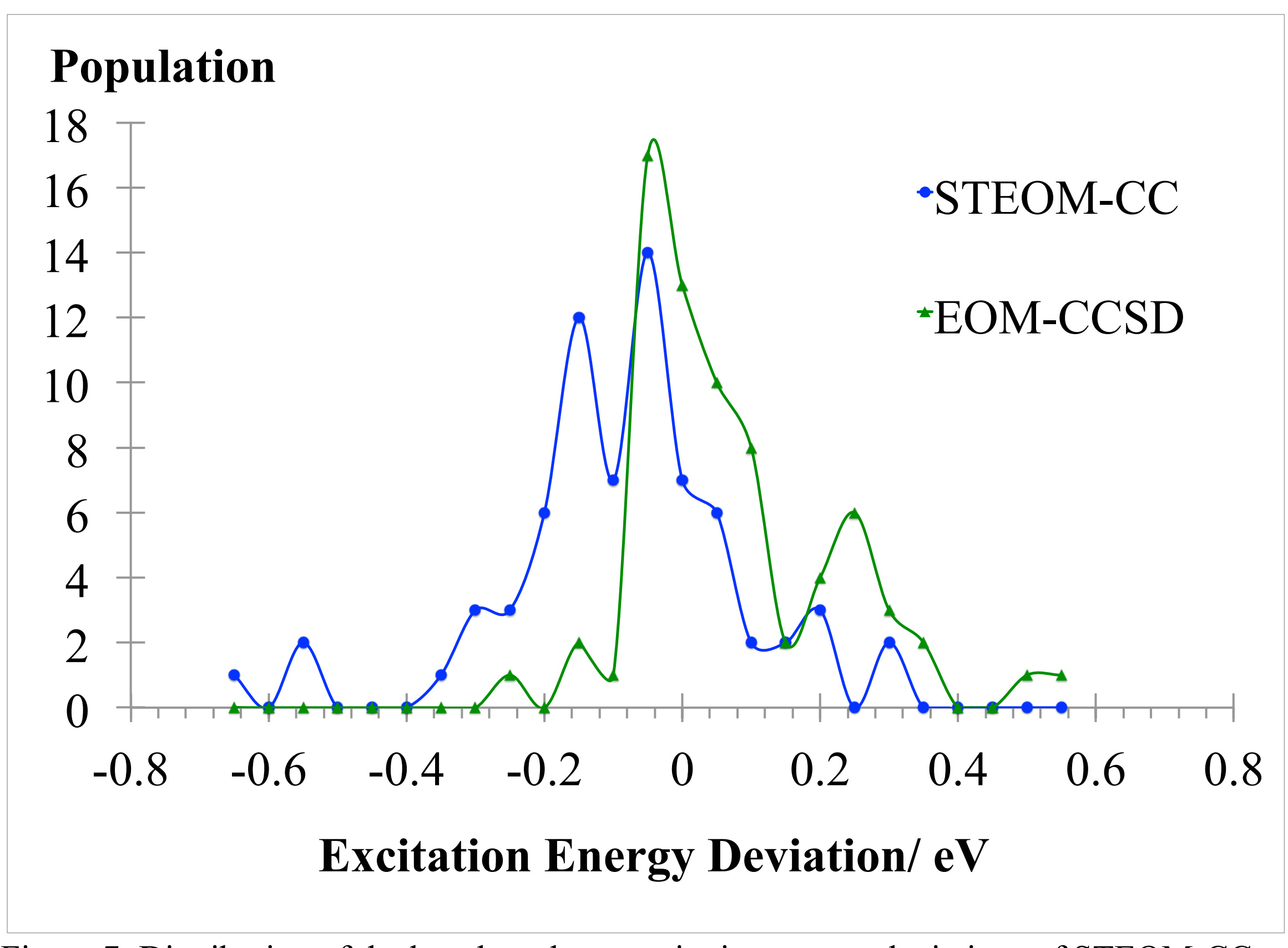

Figure 7: Distribution of the benchmark set excitation energy deviations of STEOM-CC and EOM-CCSD from EOM-CCSDT-3 for triplet states. Bin size: 0.05.

For triplet states STEOM-CC results show larger errors than EOM-CCSD (the root mean square error is ~ 0.20 eV for STEOM-CC compared to ~ 0.16 eV for EOM-CCSD). The difference is also evident from the plots. The error or deviation for the triplet states in STEOM-CC is even greater than in the case of singlet states. The improved accuracy of EOM-CCSD for triplet states is expected. It correlates to the larger value of %T1 in the CC3 calculations. The difficulties of STEOM to describe triplet states on the other hand are unexpected. In particular some quite large deviations (~0.66 eV) occur for low-lying triplet states. It is possible that a STEOM-D treatment would correct these errors, but this approach has not yet been implemented. At present we do not have a good explanation for the worse behavior of STEOM for triplet states. It is surprising we think.

### III.B.6. STEOM-CC, STEOM-D, NEVPT2 and CASPT2 against CC3 and EOM-CCSDT-3

In this section we compare STEOM-CC, STEOM-D to NEVPT2 and CASPT2 taking CC3 and EOM-CCSDT-3 as references for the comparison.

### III.B.6.1 Singlet Excitations

We present the results of the statistical analysis of the benchmarks of STEOM-CC, STEOM-D, NEVPT2 and CASPT2 compared to CC3 and EOM-CCSDT-3 in Table 9. We only include states corresponding to T1≥87% in our analysis.

| **Statistics** | **STEOM-CC** | **STEOM-D** | **NEVPT2** | **CASPT2** |
|---|---|---|---|---|
| Minimum Error | -0.219/ -0.229 | -0.323/-0.334 | -0.770/ -0.803 | -0.830/ -0.654 |
| Maximum Error | 0.159/ 0.120 | 0.100/ -0.010 | 0.540/ 0.506 | 0.250/ 0.224 |
| Mean Error | -0.019/ -0.073 | -0.067/ -0.120 | -0.066/ -0.142 | -0.225/ -0.287 |
| Mean Absolute Error | 0.065/ 0.091 | 0.085/ 0.120 | 0.258/ 0.296 | 0.241/ 0.303 |
| Root Mean Square Error | 0.081/ 0.105 | 0.118/ 0.146 | 0.318/ 0.359 | 0.291/ 0.346 |
| Standard Deviation | 0.079/ 0.076 | 0.097/ 0.084 | 0.313/ 0.333 | 0.186/ 0.196 |

Table 9: Statistical analysis of the benchmark set excitation energy deviations in eV of STEOM-CC, STEOM-D, NEVPT2 and CASPT2 from CC3 and from EOM-CCSDT-3 for singlet states. Only CC3 and EOM-CCSDT-3 states with T1≥87% are included in the analysis.

The distributions are shown below (Figures 8a and 8b).

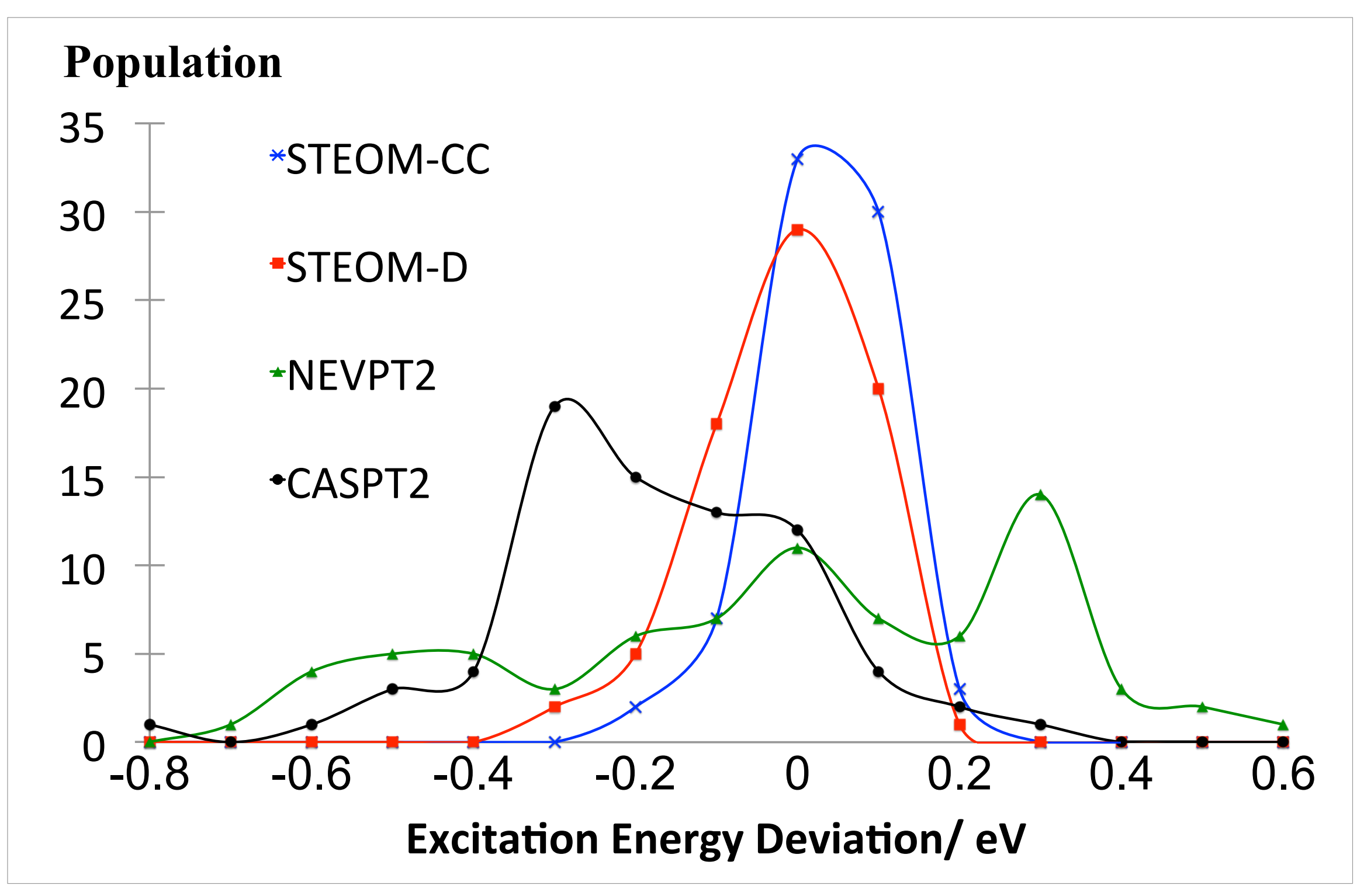

Figure 8a: Distribution of the benchmark set excitation energy deviations of STEOM-CC, STEOM-D, NEVPT2 and CASPT2 from CC3 for singlet states. Only CC3 states with T187% are included in the analysis. Bin size: 0.1.

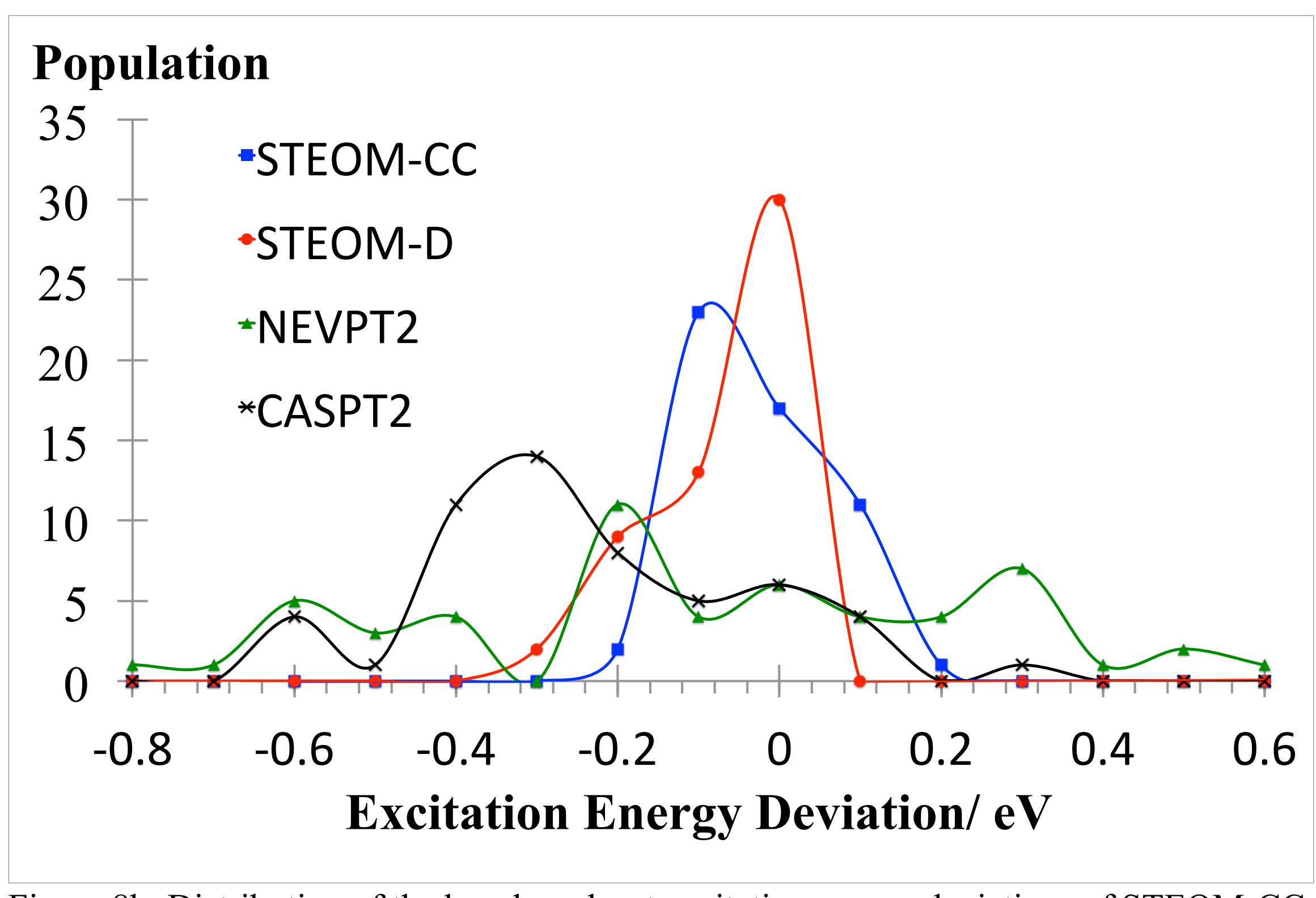

Figure 8b: Distribution of the benchmark set excitation energy deviations of STEOM-CC, STEOM-D, NEVPT2 and CASPT2 from EOM-CCSDT-3 for singlet states. Only EOM-CCSDT-3 states with T1≥87% are included in the analysis. Bin size: 0.1.

The plots and statistics show that STEOM-CC and STEOM-D are clearly superior to the reputable NEVPT2 and CASPT2 approaches when compared against CC3 and EOM-CCSDT-3. The statistical results show that STEOM-CC and STEOM-D are more accurate than NEVPT2 and CASPT2, while also the distributions show that there are generally less outliers for the STEOM methods compared to NEVPT2 and CASPT2. It is not clear whether STEOM-D is an improvement over STEOM-CC.

#### III.B.6.2 Triplet Excitations

We present the results of the statistical analysis of the benchmarks of STEOM-CC, STEOM-D, NEVPT2 and CASPT2 compared to CC3 in Table 10 and plot curves for the distribution in Figure 9.

| Statistics | STEOM-CC | EOM-CCSD | NEVPT2 | CASPT2 |
|---|---|---|---|---|
| Minimum Error | -0.662 | -0.280 | -0.820 | -0.380 |
| Maximum Error | 0.256 | 0.510 | 0.370 | 0.120 |
| Mean Error | -0.111 | 0.053 | 0.054 | -0.031 |
| Mean Absolute Error | 0.156 | 0.114 | 0.170 | 0.075 |
| Root Mean Square Error | 0.201 | 0.156 | 0.221 | 0.112 |
| Standard Deviation | 0.169 | 0.148 | 0.216 | 0.108 |

Table 10: Statistical analysis of the benchmark set excitation energy deviations in eV of STEOM-CC, STEOM-D, NEVPT2 and CASPT2 from CC3 for triplet states.

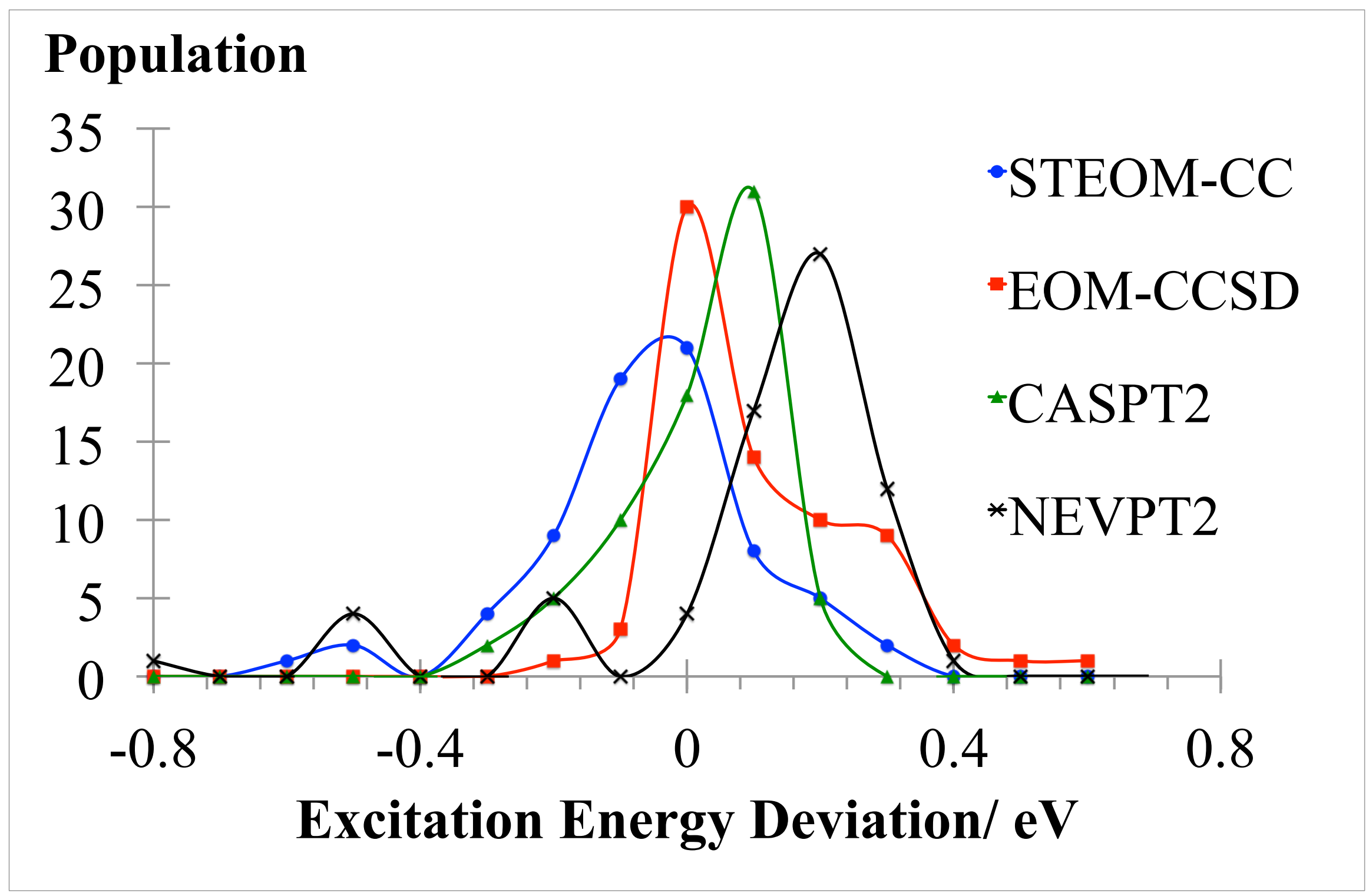

Figure 9: Distribution of the benchmark set excitation energy deviations of STEOM-CC, STEOM-D, NEVPT2 and CASPT2 from CC3 for triplet states. Bin size: 0.1.

The best method according to the distributions appear to be CASPT2 with all other methods in the plot slightly shifted to the right or to the left. STEOM-CC has a broader distribution, but is centered slightly below zero (around -0.1 eV). We hope that the STEOM-D approach can provide more accurate results for triplets. However, this is something we should attempt in the future.

### III.B.7. Doubly Excited States

To study doubly excited states, we compare CC3 states and EOM-CCSDT-3 states with relatively low %T1 and the unbiased EXT-STEOM and NEVPT2 against CASPT2. We take the CC3 T1<80% as our diagnostic for the states that we can designate as doubly excited and hence include in the statistical analysis. We present the results of the statistical analysis in Table 11.

| **Statistics** | **CC3** | **EOM-CCSDT-3** | **EXT-STEOM** | **NEVPT2** |
|---|---|---|---|---|
| Minimum Error | -0.090 | 0.056 | -0.392 | -0.590 |
| Maximum Error | 0.450 | 0.693 | 0.893 | 0.430 |
| Mean Error | 0.206 | 0.403 | 0.299 | 0.136 |
| Mean Absolute Error | 0.223 | 0.403 | 0.420 | 0.221 |
| Root Mean Square Error | 0.255 | 0.439 | 0.476 | 0.261 |
| Standard Deviation | 0.155 | 0.179 | 0.380 | 0.229 |

Table 11: Statistical analysis of the benchmark set excitation energy deviations in eV of CC3, EOM-CCSDT-3, EXT-STEOM and NEVPT2 from CASPT2 for singlet states. Only states corresponding to T1<80% in CC3 are included in the analysis.

Below is the histogram analysis for the results represented as continuous curves (Figure 10).

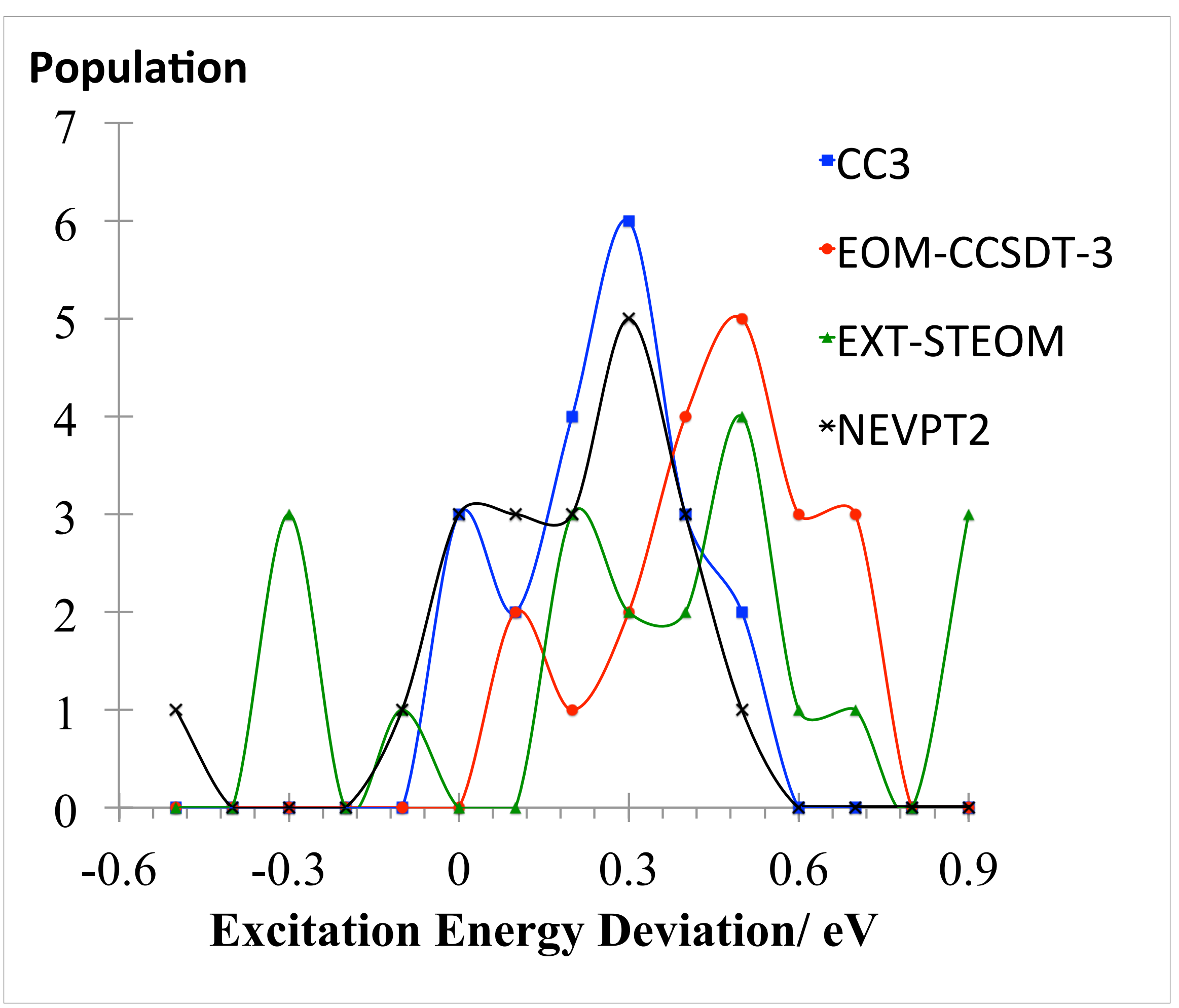

Figure 10: Distribution of the benchmark set excitation energy deviations of CC3, EOM-CCSDT-3, EXT-STEOM and NEVPT2 from CASPT2 for singlet states. Only states corresponding to CC3 T1<80% are included in the analysis. Bin size: 0.1.

We can see from the graph that CC3, EOM-CCSDT-3, EXT-STEOM, and NEVPT2 more or less follow a similar trend with two- or three-peaked distributions, while being increasing erratic in the order mentioned above. We should note that the CASPT2 approach itself is not a good reference for doubly excited states and therefore one of the few possible conclusions that can be drawn from the above results is that EXT-STEOM is not a much less accurate method than CC3, EOM-CCSDT-3 and NEVPT2 for doubly excited states. We realize that we should attempt a future benchmark against a method suitable for doubly excited states to study the effectiveness of EXT-STEOM for such states.

### IV. Concluding Remarks

The STEOM-CC approach and several of its variations have been applied to a large set of singly excited valence excited states of organic molecules. The STEOM-CC results for this large set of molecules demonstrates the robustness of the approach: while the approach in principle can break down for reasons discussed in section II, it has not for any of the states considered in this test set.
From the statistical analysis, it clearly transpires that STEOM-CC results are a significant improvement over EOM-CCSD, CASPT2, and NEVPT2 for singlet excited states. This conclusion does not hold for triplet excited states, however. The other approaches are clearly more accurate for triplets than for singlets, while STEOM can be more erratic for triplets and tends to be a bit less accurate for triplets than for singlets.
In this study only valence excited states are considered. For gas phase molecules Rydberg states are of considerable interest also. STEOM-CC and EOM-CCSD have an essential black box character that allows a convenient treatment of valence and Rydberg excited states. They are more cumbersome for CASPT2 and NEVPT2 as they require special selection of active spaces.

Regarding computational costs, STEOM-CC is clearly more cost-effective than EOM-CCSD. Compared to CASPT2 and NEVPT2 the comparison is more ambiguous. For small active spaces CASPT2 and NEVPT2 are more efficient than canonical-orbital EOM-CC and STEOM-CC approaches. Recent advances in local correlation and pair natural orbital CC approaches may affect the balance. In this context STEOM-CC may have a particular advantage as the parent state CC, IP-EOM-CC and EA-EOM-CC all naturally localize. Only at the level of the final (CIS) diagonalization step does one lose localization.

In this paper we reached some other clear conclusions. The STEOM-H (ω*) approach seems an interesting way to hermitize the transformed Hamiltonian, and is clearly superior over straightforward averaging of $G$ and $G^\dagger$. It is quite satisfactory that STEOM does not sensitively depend on the precise definition of occupied and virtual orbitals as seen in section III.B.3. Replacing the CCSD step by MBPT(2) leads to significant errors, although this approach may be still of interest in practice.
The comparison between STEOM-D and STEOM-CC did not show a clear improvement due the perturbative doubles correction. This extension may not be worthwhile. The verdict is not unambiguous however. From the discussion in section III.B.5., it appears that a particular class of excitations involving 'double bond O' groups found in amides, ketones and aldehydes produce large errors (~ 0.2 eV in STEOM-D) and affects the statistics. Further investigations are desirable. Likewise the results from EXT-STEOM are disappointing. The approach provides a respectable description of doubly excited states, but is less accurate than STEOM-CC itself for states clearly dominated by singly excited states.

In the near future we plan to present a benchmarking study considering also Rydberg and charge-transfer excited states. The STEOM-CC approach is expected to describe all states about equally well, providing a balanced approach, while not being sensitive to complications due to valence-Rydberg mixing. Such a balanced description is more of a problem to EOM-CCSD, which is more accurate for Rydberg states. Likewise CASPT2 and NEVPT2 are more problematic. Therefore, STEOM-CC might be a very attractive alternative to these fully realistic molecules in the gas phase, in particular if combined with efficient local correlation approaches.

## Acknowledgments

The original work on STEOM-CC was done in the group of Rodney J. Bartlett, and the library of ACES II subroutines developed then is still in active use today (see for example references [71, 72]). M. N. is grateful to Rodney J. Bartlett for providing such a stimulating environment, when he was a postdoctoral associate in his group (1993-1997). It is a pleasure to contribute this paper in this festschrift on the occasion of Rodney's 70th birthday and to revisit methodology conceived in those days of fruitful collaboration. This work was financially supported by a discovery grant from the Natural Sciences and Engineering Research Council of Canada (NSERC).